\documentclass[fleqn,10pt]{wlscirep}
\usepackage[utf8]{inputenc}
\usepackage[T1]{fontenc}
\usepackage{moreverb, url}
\usepackage{soul}  
\usepackage{graphicx}
\usepackage{amsmath}
\usepackage{amssymb}
\usepackage{xcolor}
\usepackage{bbold}
\usepackage{url}
\usepackage{lineno}
\usepackage{rotating}

\title{A simulation study of methods for handling disease progression in dose-finding clinical trials}

\author[1,*]{Lucie Biard}
\author[1]{Bin Cheng}
\author[2]{Gulam A. Manji}
\author[1]{Shing M. Lee}
\affil[1]{Department of Biostatistics, Mailman School of Public Health, Columbia University, New York, NY 10032, USA}
\affil[2]{Division of Medical Oncology, Columbia University Irving Medical Center, and New York Presbyterian Hospital, Herbert Irving Pavilion, New York, NY 10032, USA}

\affil[*]{lucie.biard@univ-paris-diderot.fr}

\date{6 September 2019}


\begin{abstract}
{\bf Background} 
In traditional dose-finding studies, dose-limiting toxicity (DLT) is determined within a fixed time observation window where DLT is often defined as a binary outcome. In the setting of oncology dose-finding trials, often patients in advanced stage of diseases are enrolled.  Therefore, disease progression may occur within the DLT observation window leading to treatment discontinuation and rendering the patient unevaluable for DLT assessment. As a result, additional patients have to be enrolled, increasing the sample size.

{\bf Methods} 
We propose and compare several practical methods for handling disease progression which occurs within the DLT observation window, in the context of the time-to-event continual reassessment method (TITE-CRM) which allows using partial observations. The methods differ on the way they define an evaluable patient and in the way incomplete observations are included. The methods are illustrated and contrasted in the context of a single simulated trial, and compared via simulations under various scenarios of dose-progression relationship, in the setting of advanced soft-tissue sarcoma. 

{\bf Results} In the presence of incomplete observations due to progression, the probability of correct MTD selection (PCS) decreases and the probability of
overdose selection (POS) increases. When the expected rate of progression is moderate (20 to 60\%), inclusion of incomplete observations in computations
and compensating every unevaluable patient with an additional inclusion, is warranted to maintain the PCS value in the absence of progression, while controlling for POS. When the progression rate is low ($<$20\%), the impact on PCS and POS is limited and increase of the planned sample size may be avoided. When the progression rate is expected to be above 60\%, one may consider using an alternative design or revising the definition of the toxicity observation window and/or the eligibility criteria for the trial.

{\bf Conclusions} When designing dose-finding clinical trials, it is important to consider the median progression-free survival of the population that will be
included in the study. Specific strategies should be used to limit the impact of treatment discontinuation due to progression on the operating characteristics of toxicity-based dose-finding designs.

\emph{Date: 6 September 2019}
\end{abstract}

\begin{document}

\flushbottom

\maketitle

\thispagestyle{empty}

\section*{Introduction}
Phase I dose-finding trials aim to determine a safe dose of a new drug to be used and further assessed for activity and efficacy in subsequent trials \cite{chevret_2006}.
Depending on the targeted condition or disease, and the type of expected side effects of the drug, phase I trials are performed in either healthy volunteers or patients. Based on ethical considerations, cancer dose-finding trials have usually been conducted in patients with advanced-stage disease that have grown refractory to multiple lines of treatment, due to the serious side effects of cytotoxic chemotherapies.

In a traditional dose-finding design in oncology with cytotoxic agents, several assumptions are made. First, it assumes monotone increasing dose-efficacy and dose-toxicity relationships. Second, it assumes acute severe toxicities, usually within one or two treatment cycles (4-8 weeks after treatment initiation). Third, it assumes delayed assessment of response compared to the toxicity observation window and interval drug exposure (the cytotoxic treatment is discontinued once response is definitely assessed, as opposed to chronic exposure with cytostatic agents) \cite{chevret_2006, mathijssen_2014}.

This paradigm has been challenged in recent years with the development of new classes of anti-cancer agents such as molecularly targeted agents and immunotherapy, or refinement of pre-existing treatments (\emph{e.g.} radiation therapy) \cite{mathijssen_2014}. These new treatments, some of them cytostatic, result in new frameworks for safety evaluation: plateau dose-efficacy relationship, late-onset or chronic toxicities, prolonged drug exposure, and the convergence between the safety and the efficacy time observation windows. These suggest the need for longer toxicity observation windows beyond one or two treatment cycles to capture the toxicities resulting from such prolonged treatments \cite{lee_2016, postel_2011, postel_2014}.

However, given the advanced-stage disease of patients enrolled in dose-finding cancer trials, disease progression may occur within the toxicity observation window. For ethical reasons, dose-finding study protocols usually allow handling progression by discontinuing the experimental treatment and proposing alternative therapies if available. For dose-finding studies with dichotomous or polytomous toxicity outcome observed within a pre-specified time window (i.e. 3+3 design, continual reassessment method (CRM), escalation with overdose control (EWOC), modified toxicity probability interval design (mTPI)), the occurrence of progression precludes complete safety assessment as required by the design, rendering the patient unevaluable and resulting in mandatory patient replacement, which in turn, leads to increased study duration and cost. 
Replacing patients due to progression in studies with late-onset toxicity or short median progression-free survival can be problematic and inefficient, especially in the context of rare diseases.

In this paper, we propose and evaluate approaches for handling disease progression in the context of the time-to-event CRM, which allows use of partial observations.

\section*{Motivating examples}
The present work was motivated by two clinical examples. The first is a dose-finding trial in patients with unresectable sarcoma or malignant peripheral nerve sheath tumors that was recently completed Columbia University Medical Center (NCT02584647). The goal of the trial is to identify the maximum tolerated dose (MTD) out of five doses of interest. The trial was designed using the time-to-event CRM (TITE-CRM) \cite{cheung_2000} with a sample size of 24 patients initially, revised to 18 patients during the trial, and a pre-specified target dose-limiting toxicity (DLT) probability of 0.25.  DLT was defined as any grade 3 or higher non-hematologic toxicity or grade 4 or higher hematologic toxicity in the first two cycles of treatment (i.e. 8 weeks). The TITE-CRM was selected to allow for enrollment before fully observed outcomes at the end of 8 weeks. However, given a median progression-free survival (PFS) of 6 weeks under standard of care \cite{maki_2009}, we expected almost 50\% of discontinuation due to progression by the end of the 8-week toxicity observation window. 

The second motivating example arises from a published re-analysis of dose-finding trials of bortezomib in hematologic malignancies \cite{lee_2016}. The results pointed toward an extension of the DLT observation window (to approximately 5 treatment cycles) to be able to capture and evaluate safety including non negligible late-onset toxicities. At the approved dose of 1.3mg/m$^2$, the cumulative incidence of DLT over the whole course of treatment largely exceeded the conventional target of 20-33\%. However, the median PFS was less than 5 cycles. 

In these examples, the simplest approach for handling patients who progress within a pre-specified toxicity observation window without a DLT is to render them not evaluable and enroll additional patients to replace them.  In the first motivating example, with a DLT observation window of 8 weeks and assuming a median PFS ranging from 6 weeks under standard of care to 12 weeks at the highest dose level, the increase in the number of patients due to replacements would be of at least 37\%, and potentially as high as 60\% (i.e. more than every other patient needing replacement). The operating characteristics of the design remain the same given the maintained sample size (with replacements) used for the estimation of the MTD. However, it can lead to a large number of replacements increasing trial duration and cost, and the potential bias of selection of healthier patients \cite{winther_2016}.  

In the context of the TITE-CRM, which allows for inclusion of partial information on patient follow-up during the trial, there is flexibility in the handling of early progressions occurring within the toxicity observation window, in the absence of DLT. We propose and compare three approaches for handling treatment discontinuation, notably due to disease progression (termed ``progression'' in the remaining of the paper for simplicity), in the context of the TITE-CRM and evaluate the impact on the operating characteristics of the TITE-CRM.

The Methods section introduces notations, describes three strategies to handle progression within the TITE-CRM design and presents our simulation settings. The Results section reports the results of the simulation study and concluding remarks are provided in the Discussion section.

\section*{Methods}
\subsection*{Notations}

Consider a dose-finding trial with $K$ dose-levels of interest, $d_1, \dots, d_K$, and a sample size of $N$. Dose-finding designs usually require DLT to be defined relative to a toxicity observation window, of a known duration $T$, to identify the MTD corresponding to a pre-specified target DLT probability.
Thus, DLT is often modeled as a binary variable and may occur any time within the toxicity observation window \cite{cheung_2000, chevret_2006}.

Let $U$ be the time from inclusion to DLT for a patient. DLT occurs within the toxicity window if $U \le T$. In case of early progression at time $P$ for a patient, that is, $P \le \min\{ U, T \}$, a follow-up of duration $P$ without toxicity is available for this patient, and the toxicity outcome is censored earlier than the scheduled time $T$ by $T-P$ time-units.

For simplicity, and given the framework of oncology dose-finding trials, we assume that there is no risk of loss to follow-up in this setting.

\subsection*{Strategies to handle disease progression}
Partial information due to progression events within the toxicity observation window can be handled using the TITE-CRM. This model-based approach 
through a weighted likelihood. Thus, incomplete observations due to progression can be treated as censored and included in the model likelihood, as observations without toxicity, weighted according to the available follow-up. 

Moreover, we can define a minimum follow-up threshold to render patients who progressed evaluable for toxicity assessment. Formally, let $\phi$ be a pre-specified fraction of the toxicity window $T$, where $0\leq \phi \leq 1$; a patient who progresses is deemed evaluable at the end of follow-up if $P \ge \phi T$. If $\phi=0$, all incomplete observations censored by progression are deemed evaluable. If $\phi= 1$, all incomplete observations censored by progression are deemed unevaluable and need to be replaced.  The sample size for the TITE-CRM is the number of evaluable patients. Thus, additional patients may need to be included to compensate for unevaluable patients. 

Different strategies can be implemented depending on: the threshold $\phi$ and the use of partial observations collected data for unevaluable patients with $P < \phi T$. Even though unevaluable patients will not count in the effective sample size, their partial information can be used or not in the estimation of the MTD for dose assignments. For instance, a progression, with $P < \phi T$, may occur before the enrollment of any new patient; in that case, one could argue that this unevaluable observation could be excluded. 

\noindent Three strategies were therefore considered within the TITE-CRM design. In all of them, progression without DLT was considered as censoring resulting in incomplete observation of the toxicity window. While the strategies all maintained the same number of evaluable patients, they differed in (i) the choice of $\phi$, defining an evaluable patient, (ii) the inclusion of  unevaluable patients for the estimation of the MTD. The strategies can be summarized as follows:\\

\noindent \emph{\textbf{Strategy A}}: \begin{itemize}
\item[(i)] Let $\phi=0$. All patients are evaluable regardless of time of progression. 
\end{itemize}

\noindent \emph{\textbf{Strategy B}}: \begin{itemize}
\item[(i)] Choose a $\phi>0$. 
\item[(ii)] Include all available partial follow-up time $P$ from unevaluable patients with $P < \phi T$. 

\end{itemize}

\noindent \emph{\textbf{Strategy C}}: \begin{itemize}
\item[(i)] Choose a $\phi>0$. 
\item[(ii)] Include partial information from unevaluable patients with $P < \phi T$ only if the information has been used for previous dose assignments.  
\end{itemize}
Of note, the number of evaluable patients is the same in all three strategies. In strategy A, all patients are evaluable since $\phi=0$, so the number of evaluable patients corresponds to the included patients. In strategies B and C, $\phi >0$ implies that the number of included patients may exceed the number of evaluable patients, to compensate for unevaluable patients. 

The strategies are further exemplified in a sample trial detailed in section 2.4 and illustrated in Figure \ref{fig:flow}.

\subsection*{Simulation settings}
We examined the different strategies to handle progression within the TITE-CRM design, in a simulation study in the context of the sarcoma motivating example. The design parameters were $K=5$, $N=24$, $T=8$ (corresponding to 2 cycles of treatment, with toxicity and progression assessment every week) and a target DLT probability of 0.25. The median PFS under standard of care was assumed to be 6 weeks, resulting in a progression rate of approximately 60\% by $T$.  For the estimation of the MTD, the Bayesian framework of the TITE-CRM with the empirical model for the probability of DLT by $T$ was used. The cohort size was 1 and no dose-skipping was allowed during dose escalation. The skeleton was selected using the indifference interval calibration approach. The optimal length of the indifference interval was $\delta=0.10$ based on 2000 simulations \cite{lee_2009}. To evaluate the impact of sample size, we also performed simulations with a sample size $N=18$ and its corresponding indifference interval of $\delta=0.09$.

In the design of the trial, five different scenarios of dose-toxicity relationship were evaluated, depending on the MTD true location. 
The dose-toxicity relationship was assumed to be monotone increasing, in line with a previous publication 
(see Table~\ref{tab:dosetox}) \cite{wages_2015}.
\begin{table}[h!]
\centering
\caption{Dose-toxicity relationship in simulations scenarios: true dose limiting toxicity (DLT) rate by toxicity window $T$ for each dose level, 1 to 5} \label{tab:dosetox}
\begin{tabular}{c}
  \hline
	${\rm Pr(DLT)}$ by $T$ for dose levels 1 to 5 (\%)\\
  \hline
 $(00, 01, 05, 10, {\bf 25})$\\
		 $(01, 05, 10, {\bf 25}, 40)$\\
		 $(05, 10, {\bf 25}, 40, 55)$\\
		 $(10, {\bf 25}, 40, 55, 65)$\\
		$({\bf 25}, 40, 55, 65,70)$\\
\hline
\end{tabular}
\end{table}

For each of the toxicity scenarios, we evaluated 11 scenarios of dose-progression relationships. More specifically, five of them assumed constant progression rates of 0\%, 20\%, 40\%, 60\% or 80\% by $T$ across all dose levels, corresponding to a reference situation or absence of dose effect on progression. The other six progression scenarios assumed dose-dependent progression rates: monotone decreasing (60\%, 50\%, 40\%, 30\%, 20\%) corresponding to the setting of conventional cytotoxic chemotherapy, plateau with step-decrease from 60\% to 40\% at dose-level 2, 3, 4 or 5, or U-shaped (60\%, 50\%, 40\%, 50\%, 60\%), to evaluate non strictly monotone relationships as found with new anticancer agents such as immunotherapies \cite{wages_2015, riviere_2018}.

Time to event outcomes were generated using conditional uniform models, similar to the ones used in the original TITE-CRM paper \cite{cheung_2000}. For each observation, the presence of each event within the observation window $T$ was first determined using a Bernoulli distribution with parameter corresponding to the latent cumulative incidences at $T$ described above, and if so, a time to the event was drawn from a uniform distribution on $[0, T]$. Time to toxicity and time to progression were generated independently; we then defined the observation as the time to the first event or the end of the window $T$ (in case of no event). Observations with no event or with a DLT are complete, as per the TITE-CRM design, whereas observations with a progression ($min(U, P, T)=P$) are incomplete.

Simulations were performed with a fixed accrual of 2 patients per toxicity observation window (1 per 4-week treatment cycle). Moreover, to evaluate the impact of $\phi$, simulations were performed for $\phi$ values of 0.25, 0.50 and 0.75.

The progression-handling strategies were compared based on probability of correct dose selection (PCS), probability of overdose selection (POS),  the number of additional patients enrolled in replacement of unevaluable patients and the duration of the trial.

For each scenario, 10000 replicates were performed, so that the observed PCS would be expected to lie within a 2\%-wide interval in 95\% of cases, if the true PCS of the design was 0.60. R statistical platform, version 3.4.2 and the \texttt{dfcrm} library, version 0.2-2, were used for the computations.

\subsection*{Sample trial}
\begin{figure}
\centering
  \caption{{\bf Simulated sample trial.}
    Patients flow according to the three strategies, in a trial scenario with targeted sample size $N=18$ evaluable patients, five candidate dose levels, $T=8$ weeks, $\phi = 0.5$. Enumerated segments represent patients observations, with solid lines for follow-up information included in the MTD TITE-CRM estimation, and doted lines for excluded follow-up. Patients within dashed horizontal lines received the same dose level, denoted on the y-axis. (DLT: Dose-Limiting Toxicity observed).}
\label{fig:flow} 
\includegraphics[width=5.3in]{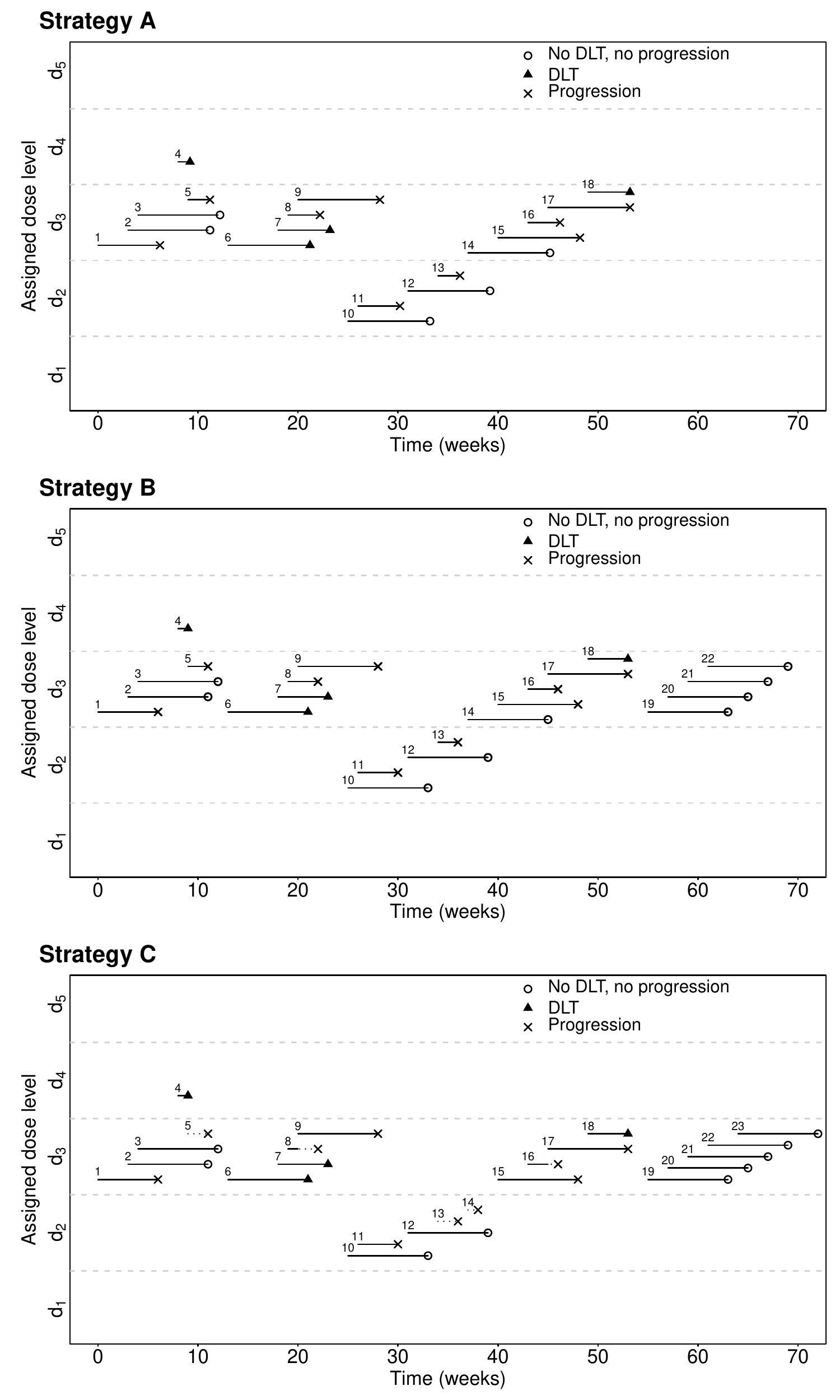}
\end{figure}

To illustrate and contrast the three strategies, we applied them to a simulated sample trial, presented in figure \ref{fig:flow}.
A trial with sample size $N=18$ was generated under the simulation scenario with a plateau dose-progression relationship with marginal progression incidences at $T$: $(60\%,60\%,40\%,40\%,40\%)$ as described above. The observation window $T$ was 8~weeks, and $\phi=0.5$ defining evaluable patients with progression in strategies B and C when $P\ge 4$~weeks.

Figure \ref{fig:flow} illustrates the differences among the strategies. Solid segments represent follow-up which is included in the estimation of the MTD  (through weights in the TITE-CRM likelihood); dotted segments correspond to follow-up which is not included. Both strategies A and B include all available information, even very short partial observations due to early progression (patients \#5, 8, 13, 16). However, these patients are considered unevaluable under strategy B and replaced, resulting in 4 additional recruitments, for a total sample size of 22. Strategy C includes partial follow up from unevaluable patients to the extent that it has been used for dose assignment computations during the trial. In other words, if a patient is deemed unevaluable, $P < \phi T$, before the next patient is included, this incomplete observation is excluded in strategy C, while included in strategies A and B (patients \#5, 13 and 14). Moreover, if the patient is deemed unevaluable ($P < \phi T$) after subsequent patient(s) has(have) been enrolled, only the information that has previously been used for dose assignment computations is included in the remaining of the trial with strategy C (e.g. patients \#8 and 16). Overall, this may result in slightly different dose assignments and total number of included patients, as illustrated in figure \ref{fig:flow}. Notably, patient \#14 is assigned to dose level~3 with strategies A and B, but, more conservatively, to dose level~2 in strategy C. Since patient \#14 progressed early when assigned to dose level 2, and before patient \#15 is included, strategy C requires one more patient compared to strategy B, in this example.

\section*{Results}
For simplicity and brevity, results from  five progression scenarios along with a reference scenario with no progression, with true MTD dose level ranging from 1 to 5, and $\phi=0.5$, are reported hereafter. Full results, including all combinations of toxicity and progression scenarios and $\phi$ parameter, are available in Supplemental material.
Tables \ref{tab:res24} and \ref{tab:res18} present results for planned sample sizes of 24 and 18 patients, respectively.

\begin{table}[h!]
\centering
\caption{Percent of correct selection (PCS), overdose selection (POS), and mean increase in sample size ($+N$ (\%)) for strategies A, B, and C, according to DLT and progression latent rates at $T$ (\%), in scenarios with 5 candidate dose levels, from 10000 simulations with $N=24$} \label{tab:res24}
{\small
\begin{tabular}{cccccccccccc}
  \hline
    & & \multicolumn{2}{c}{Strategy A}&&  \multicolumn{3}{c}{Strategy B}&&  \multicolumn{3}{c}{Strategy C}\\ \cline{3-4}\cline{6-8}\cline{10-12}
	 ${\rm Pr(DLT)}$ &${\rm Pr(Progression)}$ & PCS & POS &&   PCS& POS & $+N$(\%) &&  PCS &POS&  $+N$ (\%) \\  \hline
	$({\bf 25},40,55,65,70)$ &  $(00,00,00,00,00)$ & {\bf 68.5} & {\bf 31.5 }&  &{\bf 68.5} &{\bf 31.5} &{\bf 0 (0) }&  &{\bf 68.5} &{\bf 31.5} &{\bf 0 (0) }\\
   & $(20,20,20,20,20)$ & 64.7 & 35.3 &  & 66.0 & 34.0 & 1.9 (8) &  & 67.5 & 32.5 & 1.9 (8) \\
   & $(60,60,60,60,60)$  & 57.2 & 42.8 &  & 61.1 & 38.9 & 6.2 (26) &  & 67.4 & 32.7 & 6.5 (27) \\
   & $(60,50,40,30,20)$ & 60.5 & 39.6 &  & 62.6 & 37.5 & 5.5 (23) &  & 67.7 & 32.3 & 5.5 (23) \\
   & $(60,40,40,40,40)$ & 61.6 & 38.4 &  & 63.5 & 36.5 & 5.0 (21) &  & 68.3 & 31.7 & 5.3 (22) \\
   & $(60,60,60,60,40)$ & 57.1 & 42.9 &  & 61.0 & 39.0 & 6.2 (26) &  & 67.4 & 32.6 & 6.2 (26) \\
	   &  &  &  &  &  & &  &  &  & &  \\
$(10, {\bf 25},40,55,65)$ & $(00,00,00,00,00)$&{\bf 62.7} &{\bf 25.0} &  &{\bf 62.7} &{\bf 25.0} &{\bf 0 (0)} &  &{\bf 62.7} &{\bf 25.0} &{\bf 0 (0)}\\
& $(20,20,20,20,20)$ &59.9 & 27.5 &  & 61.6 & 26.6 & 1.9 (8) &  & 62.6 & 24.8 & 1.9 (8)\\
&$(60,60,60,60,60)$  &53.4 & 33.9 &  & 57.3 & 31.6 & 6.5 (27) &  & 58.8 & 26.6 & 6.5 (27)\\
& $(60,50,40,30,20)$  &56.0 & 30.9 &  & 60.1 & 28.8 & 4.8 (20) &  & 60.8 & 25.5 & 4.8 (20)\\
&$(60,40,40,40,40)$& 56.4 & 30.2 &  & 60.9 & 28.2 & 4.3 (18) &  & 60.4 & 25.7 & 4.3 (18) \\
&$(60,60,60,60,40)$& 53.5 & 33.9 &  & 57.4 & 31.6 & 6.5 (27) &  & 59.1 & 26.5 & 6.5 (27)\\
   &  &  &  &  &  & &  &  &  & &  \\

$(05,10, {\bf 25},40,55)$ & $(00,00,00,00,00)$ &{\bf 64.6} &{\bf 18.9} &  &{\bf 64.6} &{\bf 18.9} &{\bf 0 (0)} &  &{\bf 64.6} &{\bf 18.9} &{\bf 0 (0)} \\
  &  $(20,20,20,20,20)$& 61.8 & 21.1 &  & 63.0 & 20.7 & 1.9 (8) &  & 62.6 & 19.8 & 1.9 (8) \\
  & $(60,60,60,60,60)$ & 55.8 & 26.1 &  & 60.7 & 23.6 & 6.5 (27) &  & 61.9 & 18.4 & 6.5 (27) \\
  & $(60,50,40,30,20)$ & 59.8 & 22.4 &  & 62.8 & 21.7 & 4.1 (17) &  & 61.7 & 19.6 & 4.1 (17) \\
  & $(60,40,40,40,40)$ & 58.8 & 23.6 &  & 62.2 & 22.2 & 4.1 (17) &  & 61.6 & 20.1 & 4.1 (17) \\
  & $(60,60,60,60,40)$ & 56.0 & 26.4 &  & 59.4 & 26.3 & 6.5 (27) &  & 61.1 & 21.0 & 6.5 (27) \\
     &  &  &  &  &  & &  &  &  & &  \\

$(01, 05,10, {\bf 25},40)$ & $(00,00,00,00,00)$&{\bf 65.5} & {\bf 13.7 }&&{\bf 65.5 }&{\bf 13.7 }&{\bf 0 (0)} &&{\bf 65.5} &{\bf 13.7} &{\bf 0 (0)}\\
& $(20,20,20,20,20)$ &62.1 & 16.4 && 64.4 & 15.4 & 1.9 (8) && 64.2 & 14.3 & 1.9 (8)\\
&$(60,60,60,60,60)$  &56.1 & 20.6 && 61.3 & 18.8 & 6.5 (27) && 60.5 & 15.5 & 6.5 (27) \\
& $(60,50,40,30,20)$ &61.2 & 16.4 && 63.9 & 15.8 & 3.1 (13) && 62.8 & 14.6 & 3.1 (13)\\
& $(60,40,40,40,40)$& 60.7 & 17.4 && 63.0 & 16.8 & 4.1 (17) && 62.3 & 14.9 & 4.1 (17)\\
&$(60,60,60,60,40)$& 59.3 & 17.8 && 62.9 & 17.5 & 6.0 (25) && 61.5 & 14.3 & 6.2 (26)\\
   &  &  &  &  &  & &  &  &  & &  \\

$(00, 01, 05,10, {\bf 25})$ & $(00,00,00,00,00)$&{\bf  69.1} & n/a &  &{\bf 69.1} & n/a &{\bf 0 (0) }&&{\bf 69.1} & n/a & {\bf 0 (0)}\\
& $(20,20,20,20,20)$ &69.2 & n/a &  & 70.6 & n/a & 1.9 (8) && 69.5 & n/a & 1.9 (8)\\
&$(60,60,60,60,60)$  &67.7 & n/a &  & 73.0 & n/a & 6.7 (28) && 67.4 & n/a& 6.7 (28)\\
& $(60,50,40,30,20)$ &68.3 & n/a &  & 71.1 & n/a & 2.6 (11) && 69.2 & n/a& 2.6 (11) \\
& $(60,40,40,40,40)$& 67.8 &  n/a &  & 71.6 & n/a & 4.1 (17) && 68.2 &  n/a & 4.1 (17) \\
&$(60,60,60,60,40)$&66.4 & n/a &  & 72.2 & n/a & 5.3 (22) && 68.0 & n/a & 5.5 (23)\\

\hline
\end{tabular}
}
\end{table}

\begin{table}[h!]
\centering
\caption{Percent of correct selection (PCS), overdose selection (POS), and increase in sample size ($+N$ (\%)) for strategies A, B, and C, according to DLT and progression latent rates at $T$ (\%), in scenarios with 5 candidate dose levels, from 10000 simulations with $N=18$} \label{tab:res18}
{\small
\begin{tabular}{cccccccccccc}
  \hline
    && \multicolumn{2}{c}{Strategy A}&&  \multicolumn{3}{c}{Strategy B}&&  \multicolumn{3}{c}{Strategy C}\\ \cline{3-4}\cline{6-8}\cline{10-12}
${\rm Pr(DLT)}$ &${\rm Pr(Progression)}$& PCS & POS &&   PCS& POS & $+N$ (\%)&&  PCS &POS&  $+N$ (\%) \\  \hline
$({\bf 25},40,55,65,70)$ &   $(00,00,00,00,00)$ & {\bf 66.0} & {\bf 34.0 }&  & {\bf 66.0 }& {\bf 34.0 }& {\bf 0 (0)}&  & {\bf 66.0 }& {\bf 34.0 }& {\bf 0 (0)} \\
   &   $(20,20,20,20,20)$ & 63.4 & 36.6 &  & 64.7 & 35.3 & 1.4 (8) &  & 65.9 & 34.1 & 1.4 (8) \\
   & $(60,60,60,60,60)$ & 56.9 & 43.1 &  & 60.0 & 40.1 & 4.9 (26) &  & 64.3 & 35.7 & 4.9 (26) \\
   & $(60,50,40,30,20)$ & 59.7 & 40.3 &  & 61.4 & 38.6 & 4.0 (22) &  & 65.5 & 34.5 & 4.0 (22) \\
   &  $(60,40,40,40,40)$ & 60.2 & 39.8 &  & 62.5 & 37.6 & 3.8 (21) &  & 65.9 & 34.1 & 3.8 (21) \\
   & $(60,60,60,60,40)$ & 56.6 & 43.5 &  & 59.6 & 40.5 & 4.9 (26) &  & 64.2 & 35.8 & 4.9 (26) \\
   &  &  &  &  &  & &  &  &  & &  \\
	
$(10,{\bf 25},40,55,65)$ &  $(00,00,00,00,00)$& {\bf 57.6} & {\bf 27.0} &  & {\bf 57.6 }& {\bf 27.0 }& {\bf 0 (0) }&  &{\bf  57.6 }& {\bf 27.0} & {\bf 0 (0)}\\
  &   $(20,20,20,20,20)$& 54.5 & 29.9 &  & 55.5 & 29.1 & 1.4 (8) &  & 56.1 & 27.6 & 1.4 (8) \\
   & $(60,60,60,60,60)$ & 48.1 & 35.7 &  & 52.6 & 32.9 & 4.9 (27) &  & 53.7 & 28.5 & 4.9 (27) \\
   & $(60,50,40,30,20)$ & 50.7 & 32.9 &  & 52.8 & 31.8 & 3.6 (20) &  & 54.2 & 28.4 & 3.6 (20) \\
   & $(60,40,40,40,40)$& 50.5 & 32.8 &  & 53.8 & 30.8 & 3.2 (18) &  & 54.0 & 28.0 & 3.4 (19) \\
  &$(60,60,60,60,40)$& 47.9 & 36.0 &  & 52.3& 33.1 & 4.9 (27) &  & 53.4 & 28.8 & 4.9 (27)\\
   &  &  &  &  &  & &  &  &  & &  \\
$(05,10,{\bf 25},40,55)$ &  $(00,00,00,00,00)$&{\bf 57.7 }& {\bf 21.5 }&  & {\bf 57.7 }& {\bf 21.5} & {\bf 0 (0) }&  & {\bf 57.7} & {\bf 21.5 }& {\bf 0 (0) }\\
   &   $(20,20,20,20,20)$ & 54.5 & 24.4 &  & 55.9 & 23.9 & 1.4 (8) &  & 55.4 & 22.8 & 1.4 (8) \\
&$(60,60,60,60,60)$ & 49.5 & 28.6 &  & 53.8 & 27.1 & 4.9 (27) &  & 53.7 & 23.0 & 4.9 (27)\\
&$(60,50,40,30,20)$ & 53.7 & 24.5 &  & 56.6 & 24.2 & 3.1 (17) &  & 55.0 & 22.5 & 3.1 (17) \\
& $(60,40,40,40,40)$&53.0 & 26.0 &  & 55.1 & 25.0& 3.1 (17) &  & 54.7 & 22.4 & 3.1 (17)\\
&$(60,60,60,60,40)$&  50.5 & 27.6 &  & 54.9 & 26.1 & 4.9 (27) &  & 54.2 & 22.8 & 4.9 (27)\\
   &  &  &  &  &  & &  &  &  & &  \\
$(01, 05,10,{\bf 25},40)$ &   $(00,00,00,00,00)$ &{\bf 58.0 }& {\bf 17.8} &  &{\bf 58.0} &{\bf 17.8} &{\bf (0) }&  &{\bf 58.0 }&{\bf 17.8} & {\bf (0)} \\
   &  $(20,20,20,20,20)$& 55.8 & 19.3 &  & 57.2 & 19.5 & 1.4 (8) &  & 56.7 & 18.5 &1.4 (8) \\
   & $(60,60,60,60,60)$ & 50.3 & 23.4 &  & 55.5 & 21.3 & 5.0 28) &  & 54.2 & 18.2 & 5.0 (28) \\
   & $(60,50,40,30,20)$ & 54.3 & 19.9 &  & 57.4 & 19.4 & 2.3 (13) &  & 56.5 & 18.0 & 2.5 (14) \\
   &  $(60,40,40,40,40)$& 54.0 & 20.3 &  & 57.1 & 19.7 & 3.1 (17) &  & 56.1 & 18.0 & 3.1 (17) \\
   & $(60,60,60,60,40)$ & 51.9 & 21.0 &  & 56.3 & 20.2 & 4.5 (25) &  & 54.9 & 17.5 & 4.7 (26) \\
   &  &  &  &  &  & &  &  &  & &  \\
$(00, 01, 05,10,{\bf 25})$ &   $(00,00,00,00,00)$ &{\bf 67.8} & n/a  &  &{\bf 67.8 }& n/a &{\bf 0 (0)} &  &{\bf 67.8} & n/a & {\bf 0 (0) }\\
   &  $(20,20,20,20,20)$ & 65.5 & n/a  &  & 67.6 & n/a & 1.4 (8) &  & 66.1 & n/a  & 1.4 (8) \\
   & $(60,60,60,60,60)$ & 64.8 & n/a  &  & 69.1 & n/a  & 5.0 (28) &  & 64.1 & n/a  & 5.0 (28) \\
   & $(60,50,40,30,20)$ & 64.8 & n/a  &  & 67.8 & n/a  & 2.0 (11) &  & 65.6 & n/a  & 2.0 (11) \\
   &  $(60,40,40,40,40)$ & 65.0 & n/a  &  & 68.2 & n/a  & 3.1 (17) &  & 64.4 & n/a  & 3.1 (17) \\
   & $(60,60,60,60,40)$ & 64.0 & n/a  &  & 67.8 & n/a  & 4.1 (23) &  & 64.7 & n/a  & 4.1 (23) \\
\hline
\end{tabular}
}
\end{table}

In the absence of progression, the probability of correct dose selection (PCS) using the TITE-CRM design ranged from 62.7\% to 69.1\% for $N=24$ and 57.6\% to 67.8\% for $N=18$ in the reported scenarios, depending on the true correct MTD level. The probability of overdose selection (POS) ranged from 13.7\% to 31.5\% for $N=24$ and from 17.8\% to 34.0\% for $N=18$.

Overall, increased progression rates resulted in decreased PCS and increased POS, regardless of the strategy used and similarly in trials with $N=18$ or $24$.  However, strategy A showed greater decrease in PCS and greater increase in POS compared to strategies B and C, in all scenarios, regardless of the planned sample size.  For instance, with $N=24$ and $\phi=0.5$, PCS decreased by 9.3\%  in the scenario with true MTD being dose level 2 and 60\% constant progression rate (from 62.7\% in the absence of progression to 53.4\%), compared to 5.4\% decrease for strategy B (57.3\%) and 3.9\% decrease for strategy C (58.8\%) (Table \ref{tab:res24}).  
When comparing between strategy B and C, strategy C had better PCS, in scenarios where the true MTD was dose level 1 and 2, and slightly lower PCS when the true MTD was dose level 3 to 5 (less than 1\% difference when the true MTD was dose level 3 to 9.2\% difference in extreme settings when the true MTD was dose level 5, 80\% constant progression rate). However, at the same time, strategy C's POS was consistently lower than that of strategy B, across all scenarios and settings, even reducing POS compared to reference situations without progression when $\phi=0.75$ (Tables \ref{tab:res1875} and \ref{tab:res2475}, Supplemental material).

Regarding the impact of the dose-progression relationship, PCS was most impacted in cases of constant progression across dose levels, in comparison with decreasing, plateau or U-shaped progression rates at $T$ (Tables \ref{tab:res24} and \ref{tab:res18}, and Supplemental material).

While strategy A maintained the initial planned sample size by definition, the replacements compensating unevaluable patients with progression were very similar between strategies B and C. With $\phi=0.5$, it ranged from 8\% to 28\% of the initially planned sample size $N$, in the reported scenarios (Tables \ref{tab:res24} and \ref{tab:res18}), which corresponds to 2 to 7 supplemental patients when $N=24$.

Tables S3 and S4 (Supplemental material) present simulations results for strategies B and C when $\phi=0.25$ or $0.75$, respectively. Although similar trends to those mentioned above were observed in these settings, the gain in PCS and POS of strategies B and C relative to strategy A were diminished when $\phi=0.25$, with values of PCS and POS converging to strategy A's values. On the contrary, when $\phi=0.75$, both PCS and POS improved, outperforming the reference scenario with no progression, at the cost of a greater number of unevaluable patients and increase in the number of included patients, compared to designs with $\phi=0.5$.

\section*{Discussion}
When designing dose-finding clinical trials, it is important to consider the median progression-free survival of the population that will be included in the study if treatment will be discontinued after disease progression leading to unevaluable patients. In the setting  where a substantial number of patients are expected to progress within the toxicity observation window, consideration should be given to the use of the TITE-CRM given that it allows the flexibility of using partial information. Indeed, other methods with strictly dichotomous endpoints leave no other choice than replacing every patient who progressed. Instead, we show that the use of partial information within the TITE-CRM can reduce the final number of included patients, the duration of the study and the costs associated with both. 

This paper proposes three practical strategies to handle disease progression events within the TITE CRM, as alternatives to systematic replacement, and compared their operating characteristics via extensive simulations. Based on our simulation study, progression that occurs within the observation window can have significant impact in the operating characteristics of the TITE-CRM, as progression led to reduction in the probability of correct selection (PCS) of the true MTD, and increase in the probability of overdose selection (POS) at the end of trial in most cases, regardless of how progression cases are handled. 

As expected, we show that operating characteristics of the design can be maintained only at the cost of supplemental patients. However, we show that the definition of evaluable patients with the threshold $\phi T$ and partial replenishment of unevaluable patients only can yield acceptable performances. More precisely, the choice of strategy should be trade-off between the loss in design performances (PCS, POS, duration) and the feasibility of replacing unevaluable patients. Both strategy B and C, which allowed for the replacement of pre-defined unevaluable patients, outperformed strategy A with better PCS and lower POS. Furthermore, strategy B was more aggressive than strategy C in including all partial observations, even very short ones, and recommending doses above the MTD more often. The choice of $\phi=0.5$, the fraction parameter defining evaluable patients for toxicity assessment, relatively to the observation window, seemed to be optimal in the sense that it balanced favorably the improvement in PCS and POS with a moderate increase in sample size (and therefore trial duration) compared to strategy A (no increase in sample size).

In terms of practical guidelines, our simulation study suggests that low progression (no more than 20\%) had limited impact on PCS and POS (less than 5\% decrease in PCS and increase in POS). We would therefore recommend strategy A when the expected rate of progression is low within the DLT observation window. Thus, rather than secondarily excluding and replacing patients with progression, investigators may consider including all partial observations, as evaluable, especially in the context of rare diseases, even though some observations will be incomplete due to progression. When the expected rate of progression is moderate (from 20 to 60\%) within the desired toxicity observation window, we recommend implementing strategy C with $\phi=0.5$, to maximize PCS while controlling  POS, as a trade-off for avoiding replacement of every patient who progresses. If the expected rate of progression is high (above 60\%), the definition of the toxicity observation window and/or the eligibility criteria for the trial should be reconsidered or alternative designs, such as phase I/II designs which account for efficacy information, should be evaluated \cite{zhang_2006, yuan_2009, zang_2014, wages_2015}. 

In addition to the expected rate of progression, it is important to consider other logistical issues when designing a dose-finding trial; for example, the toxicity and efficacy assessment schedule, the criteria for treatment discontinuation and the rate of accrual.  In our motivating example, we assumed that toxicity and efficacy could be assessed within the same time frame and that the accrual rate was relatively fast (i.e. multiple patients could be accrued within the toxicity observation window). 
Thus, for reproducibility and both practical and ethical reasons, we could not systematically exclude patients with progression: part of their follow-up information might have been already used in the TITE-CRM dose-assignment computations for subsequent patient entries. However, if accrual is sufficiently slow to guarantee no more then one entry per toxicity window, patients with progression could be systematically excluded and replaced; in that case a conventional CRM design might be used instead of the TITE-CRM. On the other hand, in the presence of moderate risk of progression, as long as the accrual rate is not too slow, we argue that a TITE-CRM design in conjunction with strategy C and $\phi=0.5$ should be preferable to a CRM design (or any other design relying on complete observation of dichotomous toxicity outcome) because the former results in improved efficiency, in avoiding to replace all patients who progressed and limiting the increase in trial duration and costs. 

Finally, with anti-cancer molecularly-targeted agents (MTAs) and immunotherapy agents changing the landscape of dose-finding in oncology \cite{lee_2016, kummar_2006, letourneau_2010, postel_2011, postel_2014, chiuzan_2017}, with prolonged toxicity windows, researchers have been advocating designs including efficacy in the dose-finding process for MTAs, often relying on an optimal biological dose (OBD) instead of the MTD \cite{wages_2015, yuan_2009, zang_2014, koopmeiners_2014}.
Many of these designs require both toxicity and efficacy outcomes to be observed.  However, in oncology settings, it is common that, due to ethical concerns, patients are taken off study at progression and the experimental drug is discontinued. This prevents further assessment of DLT. At the same time, if patients experience a DLT, the drug is generally discontinued as well, preventing further assessment for efficacy. Dual assessment for toxicity and efficacy is only possible if the respective endpoint events do not prevent the complete observation one another, contrary to DLT and disease progression in our setting.  

\section*{Conclusion}
Disease progression can occur during patient follow-up in dose-finding cancer clinical trials.  However, current methods generally consider patients who progress not evaluable for toxicity, increasing both trial duration and cost. This paper proposes methods for handling disease progression in the context of the TITE-CRM, which can accommodate partial observations.   Practical strategies, relying on the inclusion of incomplete observations in the sequential MTD estimation and on patient's replacement, can limit the impact of treatment discontinuation due to disease progression. These strategies are particularly useful to handle moderate degree of progression. Thus, it is important to assess time to progression in the design of dose-finding cancer trials. Our conclusions can be extended to any cause of discontinuation, assumed independent of the toxicity endpoint. If both progression and toxicity are of interest in the dose-finding process, designs relying on both toxicity and efficacy should be considered. 

\section*{List of abbreviations}
CRM: Continual reassessment method; DLT: Dose-Limiting Toxicity; MTAs: Molecularly-targeted agents; MTD: Maximum tolerated dose; OBD: Optimal biological dose; PCS: Probability of correct dose selection; POS: Probability of overdose selection; TITE-CRM: Time-to-event continual reassessment method;

\section*{Availability of data and materials}
R code for the implemented methods is available from the authors upon request.

\section*{Competing interests}
  The authors declare that they have no competing interests.

\section*{Author's contributions}
LB, BC, SML contributed to the conception of the work, the design of the simulation study and interpreted the results. GAM provided the illustrative example scenario in sarcoma. LB, BC and SML drafted and edited the manuscript. All authors (LB, BC, GAM, SML) read and approved the final manuscript.
\section*{Acknowledgements}
Shing M. Lee was supported by the American Cancer Society (grant number MRSG-13-146-01-CPHPS).

\bibliography{biblio}

\section*{Supplementary material}
The following tables report the percent of correct selection (PCS), overdose selection (POS) and average number of unevaluable patients resulting in increasing the sample size ($+N$ (\%)) for strategies A, B and C, according to the dose limiting toxicity (DLT) and progression latent rates within the toxicity observation window $T$ (\%), in scenarios with 5 candidate dose levels and 2 patients-per-toxicity window accrual rate, from 10000 simulations: 
\begin{itemize}
\item[-] Table \ref{tab:res18_supp}: $N=18$, strategy A, and strategies B and C with $\phi=0.5$,
\item[-] Table \ref{tab:res24_supp}: $N=24$, strategy A, and strategies B and C with $\phi=0.5$,
\item[-] Table \ref{tab:res1825}: $N=18$, strategies B and C with $\phi=0.25$,
\item[-] Table \ref{tab:res2425}: $N=24$, strategies B and C with  $\phi=0.25$,
\item[-] Table \ref{tab:res1875}: $N=18$, strategies B and C with  $\phi=0.75$,
\item[-] Table \ref{tab:res2475}: $N=24$, strategies B and C with  $\phi=0.75$.
\end{itemize}

\begin{table}[ht!]
\centering
{\small
\caption{Percent of correct selection (PCS), overdose selection (POS) and increase in sample size ($+N$ (\%)) for strategies A, B and C , according to dose limiting toxicity (DLT) and progression latent rates at $T$ (\%), in scenarios with 5 candidate dose levels, from 10000 simulations with $N=18$, $\phi=0.5$ and 2 patients-per-observation window accrual rate} \label{tab:res18_supp}
\begin{tabular}{llcccccccccc}
  \hline
    && \multicolumn{2}{c}{Strategy A}&&  \multicolumn{3}{c}{Strategy B}&&  \multicolumn{3}{c}{Strategy C}\\ \cline{3-4}\cline{6-8}\cline{10-12} 
	 Pr(DLT)&Pr(Progression) &  PCS & POS && PCS & POS &$+N$ (\%) &&  PCS &POS & $+N$ (\%) \\  \hline
$(\boldsymbol{25},40,55,65,70)$ &   $(00,00,00,00,00)$ &{\bf 66.0} &{\bf 34.0} && {\bf 66.0 }&{\bf 34.0} & {\bf 0 (0)} &&{\bf 66.0} &{\bf 34.0} &{\bf 0 (0)}\\ 
   &  $(20,20,20,20,20)$ & 63.4 & 36.6 && 64.7 & 35.3 & 1.4 (8) && 65.9 & 34.1 & 1.4 (8) \\ 
	   &  $(40,40,40,40,40)$ & 60.3 & 39.7&&  62.8 & 37.2 & 2.9 (16) && 65.5 & 34.5 & 2.9 (16) \\ 
   & $(60,60,60,60,60)$ & 56.9 & 43.1 && 60.0 & 40.1& 4.7 (26) && 64.3 & 35.7 & 4.7 (26) \\ 
	   & $(80,80,80,80,80)$ &51.9 & 48.2 && 56.1 & 43.9 & 6.8 (38) && 64.0 & 36.0 & 7.0 (39)\\ 
   & $(60,50,40,30,20)$ & 59.7 & 40.3 && 61.4 & 38.6 & 4.0 (22) && 65.5 & 34.5 & 4.0 (22) \\ 
   &  $(60,40,40,40,40)$ & 60.2 & 39.8 && 62.5 & 37.6 & 3.8 (21) && 65.9 & 34.1 & 3.8 (21) \\ 
   &  $(60,60,40,40,40)$ & 57.7 & 42.3 && 60.8 & 39.2 & 4.3 (24) && 64.4 & 35.7 &4.3 (24) \\ 
   &  $(60,60,60,40,40)$ &  57.1 & 42.9 && 60.0 & 40.0 & 4.7 (26) && 64.3 & 35.7 & 4.7 (26) \\ 
   & $(60,60,60,60,40)$ & 56.6 & 43.5 && 59.6 & 40.5 & 4.7 (26) && 64.2 & 35.8 & 4.7 (26) \\ 
	&  $(60,50,40,50,60)$ &  58.5 & 41.5 && 61.1 & 38.9 & 4.1 (23) && 65.0 & 35.0 & 4.1 (23) \\
	& & & && & & && & & \\

$(10, \boldsymbol{25},40,55,65)$  &  $(00,00,00,00,00)$&{\bf 57.6} &{\bf 27.0 }&  &{\bf 57.6} &{\bf 27.0 }& {\bf 0 (0)} &  &{\bf 57.6} &{\bf 27.0} & {\bf 0 (0)}\\  
  &  $(20,20,20,20,20)$ & 54.5 & 29.9 && 55.5 & 29.1 & 1.4 (8) && 56.1 & 27.6 & 1.4 (8) \\ 
	&  $(40,40,40,40,40)$ & 51.2 & 32.9 && 53.5 & 31.0 & 2.9 (16) && 54.7 & 28.3 & 3.1 (17) \\ 
   & $(60,60,60,60,60)$ & 48.1 & 35.7 && 52.6 & 32.9 & 4.9 (27) && 53.7 & 28.5 & 4.9 (27) \\ 
	 & $(80,80,80,80,80)$ & 44.9 & 39.2 && 49.8 & 36.2 & 7.0 (39) && 52.1 & 28.9 & 7.2 (40) \\ 
   & $(60,50,40,30,20)$ & 50.7 & 32.9 && 52.8 & 31.8 & 3.6 (20) && 54.2 & 28.4 & 3.6 (20) \\ 
  	& $(60,40,40,40,40)$& 50.5 & 32.8 && 53.8 & 30.8 & 3.2 (18) && 54.0 & 28.0 & 3.4 (19) \\ 
     &  $(60,60,40,40,40)$ &51.1 & 33.1 && 54.0 & 31.5 & 4.0 (22) && 54.0 & 28.5 & 4.1 (23) \\ 
   &  $(60,60,60,40,40)$ & 48.7 & 35.0 &  & 52.5 & 32.8 & 4.7 (26) && 53.4 & 28.8 & 4.7 (26) \\ 
	&$(60,60,60,60,40)$& 47.9 & 36.0&  & 52.3 & 33.1 & 4.9 (27) && 53.4 & 28.8 & 4.9 (27)\\
&  $(60,50,40,50,60)$ &  49.9 & 33.9 && 52.9 & 31.7 & 3.8 (21) && 53.8 & 28.5 & 3.8 (21)  \\
	& & & && & & && & & \\

$(05,10, \boldsymbol{25},40,55)$ &  $(00,00,00,00,00)$&{\bf 57.7} &{\bf 21.5} &&{\bf 57.7 }&{\bf 21.5} &{\bf 0 (0)} &&{\bf 57.7} &{\bf 21.5} &{\bf 0 (0)} \\  
 &  $(20,20,20,20,20)$ & 54.5 & 24.4 && 55.9 & 23.9 &1.4 (8) && 55.4 & 22.8 & 1.4 (8) \\ 
&  $(40,40,40,40,40)$ & 52.5 & 26.2 && 55.1 & 25.4 & 3.1 (17) && 54.7 & 23.0 & 3.1 (17) \\ 
&$(60,60,60,60,60)$ & 49.5 & 28.6 && 53.8 & 27.0 & 4.9 (27) && 53.7 & 23.0 & 4.9 (27)\\
 & $(80,80,80,80,80)$ & 46.8 & 30.8 && 53.1 & 28.7 & 7.2 (40) && 52.6 & 22.6 & 7.2 (40)\\ 
&$(60,50,40,30,20)$ & 53.7 & 24.5 && 56.6 & 24.2 &3.1 (17) && 55.0 & 22.5 & 3.1 (17) \\  
& $(60,40,40,40,40)$&53.0 & 26.0 && 55.1 & 25.0 & 3.1 (17) && 54.7 & 22.4 & 3.1 (17)\\  
&  $(60,60,40,40,40)$ & 51.1 & 26.2 && 54.3 & 25.4 & 3.6 (20)&& 54.0 & 22.7 & 3.6 (20)\\ 
&  $(60,60,60,40,40)$ & 52.2 & 25.6 && 55.7 & 24.9 & 4.3 (24) && 54.9 & 21.6 & 4.5 (25)\\ 
&$(60,60,60,60,40)$&  50.5 & 27.6 && 54.9 & 26.1 & 4.9 (27) && 54.2 & 22.8 & 4.9 (27)\\
&  $(60,50,40,50,60)$ & 51.2 & 27.0  &&  54.4 & 26.1 & 3.6 (20) && 54.0 & 23.1 & 3.6 (20) \\
	& & & && & & && & & \\

$(01, 05,10, \boldsymbol{25},40)$ & $(00,00,00,00,00)$ &{\bf 58.0} &{\bf 17.8} &&{\bf 58.0} &{\bf 17.8} & {\bf 0 (0)} &&{\bf 58.0}&{\bf 17.8} &{\bf 0 (0)}\\ 
   &  $(20,20,20,20,20)$ & 55.8 & 19.3 && 57.2 & 19.5 & 1.4 (8) && 56.7 & 18.5 &1.4 (8) \\ 
	&  $(40,40,40,40,40)$ & 53.4 & 21.3 && 56.4 & 20.5 & 3.1 (17) && 56.0 & 18.3 & 3.1 (17) \\ 
   & $(60,60,60,60,60)$ & 50.3 & 23.4 && 55.5 & 21.3 & 5.0 (28) && 54.2 & 18.2 & 5.0 (28) \\
	 & $(80,80,80,80,80)$ &  47.6 & 24.9 && 53.3 & 23.6 & 7.2 (40) && 52.1 & 18.1 & 7.2 (40) \\ 
   & $(60,50,40,30,20)$ & 54.3 & 19.9 && 57.4 & 19.4 & 2.3 (13) && 56.5 & 18.0 & 2.5 (14) \\ 
   &  $(60,40,40,40,40)$& 54.0 & 20.3 && 57.1 & 19.7 & 3.1 (17) && 56.1 & 18.0 & 3.1 (17) \\ 
	   &  $(60,60,40,40,40)$ &  52.9 & 21.3 && 56.5 & 20.3 & 3.1 (17) && 55.4 & 18.6 & 3.2 (18) \\ 
   &  $(60,60,60,40,40)$ & 51.3 & 21.2 && 56.3 & 20.1 & 3.8 (21) && 54.8 & 17.8 & 3.8 (21) \\ 
   & $(60,60,60,60,40)$ & 51.9 & 21.0 && 56.3 & 20.2 & 4.5 (25) && 54.9 & 17.5 & 4.7 (26)\\ 
&  $(60,50,40,50,60)$ & 50.9 & 23.3 && 55.3 & 21.5 & 3.8 (21) && 55.4 & 18.3 & 3.8 (21) \\
	& & & && & & && & & \\

$(00, 01, 05,10, \boldsymbol{25})$ & $(00,00,00,00,00)$ &{\bf 67.8} & n/a &&{\bf 67.8} & n/a &{\bf 0 (0)} &  &{\bf 67.8} & n/a &{\bf 0 (0)} \\ 
   &  $(20,20,20,20,20)$ & 65.5 & n/a && 67.6 & n/a & 1.4 (8) && 66.1 & n/a & 1.4 (8)\\ 
	&  $(40,40,40,40,40)$ & 64.9& n/a &&  67.7 & n/a & 3.1 (17) && 64.8 & n/a & 3.1 (17) \\ 
   & $(60,60,60,60,60)$ & 64.8 & n/a && 69.1 & n/a & 5.0 (28) && 64.1 & n/a & 5.0 (28) \\
	 & $(80,80,80,80,80)$ &  63.8 & n/a && 69.3 & n/a & 7.4 (41) && 63.0 & n/a & 7.4 (41)\\ 
   & $(60,50,40,30,20)$ & 64.8 & n/a && 67.8 & n/a & 2.0 (11) && 65.6 & n/a & 2.0 (11)\\ 
   &  $(60,40,40,40,40)$ & 65.0 & n/a && 68.2 & n/a & 3.1 (17) && 64.4 & n/a & 3.1 (17)\\
	  & $(60,60,40,40,40)$ & 64.7 & n/a && 67.8 & n/a & 3.1 (17) && 64.5 & n/a & 3.1 (17)\\ 
   &  $(60,60,60,40,40)$ & 65.3 & n/a && 68.4 & n/a & 3.4 (19) && 65.2&  n/a & 3.4 (19)\\ 
   & $(60,60,60,60,40)$ & 64.0 & n/a && 67.8 & n/a  &4.1 (23) && 64.7 & n/a & 4.1 (23) \\ 
	&  $(60,50,40,50,60)$ &  64.3 & n/a && 67.7 &  n/a & 4.3 (24) && 63.5 & n/a & 4.3 (24)\\
\hline
\end{tabular}
}
\end{table}

\clearpage

\begin{table}[ht!]
\centering
{\small
\caption{Percent of correct selection (PCS), overdose selection (POS) and increase in sample size ($+N$ (\%)) for strategies A, B and C, according to dose limiting toxicity (DLT) and progression latent rates at $T$ (\%), in scenarios with 5 candidate dose levels, from 10000 simulations with $N=24$, $\phi=0.5$ and 2 patients-per-observation window accrual rate} \label{tab:res24_supp}
\begin{tabular}{llcccccccccc}
  \hline
    && \multicolumn{2}{c}{Strategy A}&&  \multicolumn{3}{c}{Strategy B}&&  \multicolumn{3}{c}{Strategy C}\\ \cline{3-4}\cline{6-8}\cline{10-12} 
Pr(DLT)&Pr(Progression) &  PCS & POS && PCS & POS &$+N$ (\%) &&  PCS &POS & $+N$ (\%) \\  \hline
	$(\boldsymbol{25},40,55,65,70)$ &  $(00,00,00,00,00)$ &{\bf 68.5} &{\bf 31.5} &  &{\bf 68.5}&{\bf 31.5} &{\bf 0 (0)}&  & {\bf 68.5 }&{\bf 31.5} &{\bf 0 (0)}\\ 
   & $(20,20,20,20,20)$ & 64.7 & 35.3 &  & 66.0 & 34.0 & 1.9 (8)&  & 67.5 & 32.5 & 1.9 (8)\\ 
	&  $(40,40,40,40,40)$ &  61.5 & 38.5 &  & 64.0 & 36.0 & 3.8 (16) &  & 67.1 & 32.9 & 3.8 (16)\\ 
   & $(60,60,60,60,60)$  & 57.2 & 42.8 &  & 61.1 & 38.9 & 6.2 (26) &  & 67.4 & 32.7 & 6.5 (27) \\ 
	 & $(80,80,80,80,80)$ & 52.5 & 47.5 &  & 57.5 & 42.5 & 9.1 (38) &  & 66.8 & 33.2 & 9.4 (39)\\ 
   & $(60,50,40,30,20)$ & 60.5 & 39.6 &  & 62.6 & 37.5 & 5.5 (23) &  & 67.7 & 32.3 & 5.5 (23) \\ 
   & $(60,40,40,40,40)$ & 61.6 & 38.4 &  & 63.5 & 36.5 & 5.0 (21) &  & 68.3 & 31.7 & 5.3 (22) \\ 
	 & $(60,60,40,40,40)$ & 58.9 & 41.1 &  & 62.4 & 37.6 & 6.0 (25) &  & 67.6 & 32.4 & 6.0 (25)\\ 
   & $(60,60,60,40,40)$ &  57.6 & 42.4 &  & 61.4 & 38.6 & 6.2 (26) &  & 67.0 & 33.0 & 6.2 (26)\\
   & $(60,60,60,60,40)$ & 57.1 & 42.9 &  & 61.0 & 39.0 & 6.2 (26) &  & 67.4 & 32.6 &6.2 (26)\\ 
	 & $(60,50,40,50,60)$ & 59.9 & 40.1 &  & 62.6 & 37.4 & 5.5 (23) &  & 67.4 & 32.6 & 5.5 (23) \\
	&   &  &  &&  & & &  &  &  & \\

$(10, \boldsymbol{25},40,55,65)$  & $(00,00,00,00,00)$&{\bf 62.7} &{\bf 25.0 }&  &{\bf 62.7 }&{\bf 25.0 }&{\bf 0 (0)}&  &{\bf 62.7} &{\bf 25.0} &{\bf 0 (0)}\\  
&$(20,20,20,20,20)$ &59.9 & 27.5 &  & 61.6 & 26.6 & 1.9 (8)&  & 62.6 & 24.8 &1.9 (8)\\  
&  $(40,40,40,40,40)$ &56.1 & 31.0 &  & 60.0 & 28.6 &3.8 (16) &  & 60.4 & 25.7 & 3.8 (16) \\ 
&$(60,60,60,60,60)$  &53.4 & 33.9 &  & 57.3 & 31.6 &6.5 (27)&  & 58.8 & 26.6 & 6.5 (27)\\
& $(80,80,80,80,80)$ &49.0 & 38.4 &  & 54.8 & 35.0 &9.6 (40)&  & 57.7 & 26.9 & 9.6 (40)  \\ 
& $(60,50,40,30,20)$  &56.0 & 30.9 &  & 60.1 & 28.8 & 4.8 (20) &  & 60.8 & 25.5 & 4.8 (20)\\  
&$(60,40,40,40,40)$& 56.4 & 30.2 &  & 60.9 & 28.2 & 4.3 (18) &  & 60.4 & 25.7 & 4.3 (18) \\ 
& $(60,60,40,40,40)$ & 55.7 & 31.5 &  & 59.2 & 29.9 & 5.5 (23) &  & 60.0 & 26.4 & 5.5 (23)\\ 
&  $(60,60,60,40,40)$ & 54.2 & 33.0 &  & 58.2 & 30.7 & 26.2 (26)&  & 59.3 & 26.0 &6.2 (26)\\ 
&$(60,60,60,60,40)$& 53.5 & 33.9 &  & 57.4 & 31.6 &6.5 (27) &  & 59.1 & 26.5 &6.5 (27)\\
&  $(60,50,40,50,60)$ & 56.1 & 30.7 &  & 60.0 & 28.9 & 5.0 (21) &  & 60.1 & 25.8 & 5.0 (21)  \\
	&   &  &  &&  & & &  &  &  & \\

$(05,10, \boldsymbol{25},40,55)$ & $(00,00,00,00,00)$ &{\bf 64.6} &{\bf 18.9} &  &{\bf 64.6 }&{\bf 18.9 }&{\bf 0 (0)} &  &{\bf 64.6} &{\bf 18.9}&{\bf 0 (0)} \\ 
  &  $(20,20,20,20,20)$& 61.9 & 21.1 &  & 63.0 & 20.7 &1.9 (8)&  & 62.6 & 19.8 & 1.9 (8)\\ 
	&  $(40,40,40,40,40)$ & 58.9 & 23.8 &  & 62.0 & 22.5 & 4.1 (17) && 62.1 & 19.9 & 4.1 (17)\\ 
  & $(60,60,60,60,60)$ & 55.8 & 26.1 &  & 60.7 & 23.6 & 6.5 (27)&& 61.9 & 18.4 &6.5 (27) \\
	 & $(80,80,80,80,80)$ &53.8 & 28.5 &  & 59.5 & 26.3 & 9.6 (40) && 59.4 & 19.7 &9.6 (40)\\ 
  & $(60,50,40,30,20)$ & 59.8 & 22.4 &  & 62.8 & 21.7 & 4.1 (17) && 61.7 & 19.6 & 4.1 (17) \\ 
  & $(60,40,40,40,40)$ & 58.8 & 23.6 &  & 62.2 & 22.2 & 4.1 (17) && 61.6 & 20.1 & 4.1 (17) \\ 
		   &  $(60,60,40,40,40)$ & 57.8 & 24.0 &  & 61.7 & 22.6 & 4.6 (19) && 60.9 & 20.0 & 4.6 (19) \\ 
   &  $(60,60,60,40,40)$ & 58.2 & 23.7 &  & 62.0 & 22.6 & 6.0 (25) &  & 61.0 & 19.4 & 6.0 (25)  \\
  & $(60,60,60,60,40)$ & 56.0 & 26.4 &  & 59.4 & 26.3 &6.5 (27)&  & 61.1 & 20.9 & 6.5 (27)\\ 
  &  $(60,50,40,50,60)$ & 57.6 & 24.7 &  & 61.8 & 23.2 & 4.8 (20) &  & 61.3 & 20.5 & 4.8 (20) \\
	&   &  &  &&  & & &  &  &  & \\

$(01, 05,10, \boldsymbol{25},40)$ & $(00,00,00,00,00)$&{\bf 65.5} &{\bf 13.7} &&{\bf 65.5 }&{\bf 13.7}&{\bf 0 (0)} &&{\bf 65.5}&{\bf 13.7}&{\bf 0 (0)}\\  
&$(20,20,20,20,20)$ &62.1 & 16.4 && 64.4 & 15.4 & 1.9 (8)&& 64.2 & 14.3 & 1.9 (8)\\ 
&  $(40,40,40,40,40)$ & 59.6 & 18.3 &  & 62.9 & 17.1 & 4.1 (17) &  & 61.8 & 15.1 & 4.1 (17) \\  
&$(60,60,60,60,60)$  &56.1 & 20.6 && 61.3 & 18.8 & 6.5 (27) && 60.5 & 15.5 &  6.5 (27) \\
 & $(80,80,80,80,80)$ &53.9 & 22.1 &  & 59.8 & 20.6 &9.6 (40)&  & 58 & 14.8 &9.6 (40)\\ 
& $(60,50,40,30,20)$ &61.2 & 16.4 && 63.9 & 15.8 & 3.1 (13) && 62.8 & 14.6 & 3.1 (13)\\  
& $(60,40,40,40,40)$& 60.7 & 17.4 && 63.0 & 16.8 & 4.1 (17) && 62.3 & 14.9 & 4.1 (17)\\
	   &  $(60,60,40,40,40)$ &  59.6 & 18.2 &  & 63.2 & 17.1 & 4.1 (17) &  & 61.7 & 15.5 & 4.1 (17) \\ 
   &  $(60,60,60,40,40)$ & 58.6 & 18.0 &  & 63.0 & 16.9 & 5.0 (21) &  & 61.7 & 15.0 & 5.0 (21) \\  
&$(60,60,60,60,40)$& 59.3 & 17.8 && 62.9 & 17.5 & 6.0 (25) && 61.5 & 14.3 &6.2 (26)\\
&  $(60,50,40,50,60)$ & 58.6 & 19.6 &  & 62.6 & 17.8 & 5.0 (21) &  & 62.2 & 14.7 & 5.0 (21) \\
	&   &  &  &&  & & &  &  &  & \\

$(00, 01, 05,10, \boldsymbol{25})$ & $(00,00,00,00,00)$&  {\bf 69.1 }& n/a &  &{\bf 69.1} & n/a &{\bf 0 (0)}&&{\bf 69.1} & n/a &{\bf 0 (0)}\\  
&$(20,20,20,20,20)$ &69.2 & n/a &  & 70.6 & n/a & 1.9 (8)&& 69.5 & n/a & 1.9 (8)\\ 
&  $(40,40,40,40,40)$ &  67.7 &  n/a &  & 71.1 &  n/a & 4.1 (17) &  & 68.5 &  n/a & 4.1 (17) \\  
&$(60,60,60,60,60)$  &67.7 & n/a &  & 73.0 & n/a & 6.7 (28) && 67.4 & n/a& 6.7 (28)\\
 & $(80,80,80,80,80)$ & 66.8 & n/a &  & 73.6 & n/a & 9.8 (41) &  & 65.5 & n/a &9.8 (41) \\ 
& $(60,50,40,30,20)$ &68.3 & n/a &  & 71.1 & n/a & 2.6 (11) && 69.2 & n/a& 2.6 (11)\\  
& $(60,40,40,40,40)$& 67.8 &  n/a &  & 71.6 & n/a & 4.1 (17) && 68.2&  n/a & 4.1 (17) \\ 
	   &  $(60,60,40,40,40)$ & 67.9 &  n/a&  & 71.3 &  n/a & 4.1 (17) &  & 68.4 &  n/a & 4.1 (17)  \\ 
   &  $(60,60,60,40,40)$ &  68.4 & n/a &  & 71.8 & n/a & 4.6 (19)&  & 68.5 & n/a & 4.6 (19)\\ 
&$(60,60,60,60,40)$&66.4 & n/a &  & 72.2 & n/a & 5.3 (22) && 68.0 & n/a & 5.5 (23)\\
&  $(60,50,40,50,60)$ & 67.5 &  n/a &  & 71.8 &  n/a & 5.8 (24) &  & 67.0 &  n/a & 5.8 (24) \\

\hline
\end{tabular}
}
\end{table}

\begin{table}[ht!]
\centering
{\small

\caption{Percent of correct selection (PCS), overdose selection (POS) and increase in sample size ($+N$ (\%)) for strategies A, B and C, according to dose limiting toxicity (DLT) and progression latent rates at $T$ (\%), in scenarios with 5 candidate dose levels, from 10000 simulations with $N=18$, $\phi=0.25$ and 2 patients-per-observation window accrual rate} \label{tab:res1825}
\begin{tabular}{llccccccc}
  \hline
    &&  \multicolumn{3}{c}{Strategy B}&&  \multicolumn{3}{c}{Strategy C}\\ \cline{3-5}\cline{7-9}
Pr(DLT)&Pr(Progression) &  PCS & POS & $+N$ (\%) &&  PCS &POS & $+N$ (\%) \\  \hline
$(\boldsymbol{25},40,55,65,70)$ & $(00,00,00,00,00)$ &{\bf 66.0} &{\bf 34.0} &{\bf 0 (0)}& &{\bf 66.0} &{\bf 34.0} & {\bf 0 (0)}\\ 
   &  $(20,20,20,20,20)$ & 63.9 & 36.1 & 0.5 (3) & & 64.2 & 35.8 & 0.4 (2) \\ 
	   &  $(40,40,40,40,40)$ & 61.0 & 39.0 & 0.9 (5) & & 61.5 & 38.5 &0.9 (5)\\ 
   & $(60,60,60,60,60)$ &57.7 & 42.3 & 1.4 (8) &  & 58.4 & 41.6 & 1.4 (8)\\ 
	   & $(80,80,80,80,80)$ &53.4 & 46.6 & 2.0 (11) &  & 55.0 & 45.0 & 2.0 (11)\\ 
   & $(60,50,40,30,20)$ & 60.2 & 39.8 & 1.3 (7) & & 60.9 & 39.1 & 1.3 (7)\\ 
   &  $(60,40,40,40,40)$ & 61.1 & 38.9 & 1.1 (6) & & 61.7 & 38.3 & 1.1 (6)\\ 
   &  $(60,60,40,40,40)$ & 58.4 & 41.6 & 1.3 (7)& & 59.0 & 41.0 & 1.3 (7)\\ 
   &  $(60,60,60,40,40)$ & 58.2 & 41.8 & 1.4 (8)& & 59.0 & 41.0 & 1.4 (8)\\ 
   & $(60,60,60,60,40)$ & 57.6 & 42.4 & 1.4 (8)& & 58.2 & 41.8 & 1.4 (8)\\ 
	&  $(60,50,40,50,60)$ & 59.5 & 40.5 & 1.3 (7)& & 60.3 & 39.7 & 1.3 (7)\\
		&   &  &  &&  & & &    \\

$(10, \boldsymbol{25},40,55,65)$  &  $(00,00,00,00,00)$&{\bf 57.6} &{\bf 27.0} &{\bf 0 (0)} &  &{\bf 57.6} &{\bf 27.0} & {\bf 0 (0)}\\  
   &  $(20,20,20,20,20)$ &54.9 & 29.6 & 0.4 (2)&  & 55.0 & 29.4 & 0.4 (2)\\ 
	   &  $(40,40,40,40,40)$ & 52.1 & 32.1 &0.9 (5)&  & 52.5 & 31.6 &0.9 (5)\\ 
   & $(60,60,60,60,60)$ &49.5 & 34.7 & 1.4 (8)&  & 49.8 & 33.7 &  1.4 (8)\\ 
	   & $(80,80,80,80,80)$ & 46.3 & 38.6 & 2.0 (11)&  & 46.2 & 37.6 & 2.0 (11)\\ 
   & $(60,50,40,30,20)$ & 51.5 & 32.2 & 1.1 (6)&  & 51.6 & 31.8 & 1.1 (6)\\ 
   &  $(60,40,40,40,40)$ &51.7 & 31.9 & 1.1 (6)&  & 51.9 & 31.3 & 1.1 (6)\\ 
   &  $(60,60,40,40,40)$ & 51.8 & 32.4 & 1.3 (7)&  & 51.7 & 31.8 & 1.3 (7)\\ 
   &  $(60,60,60,40,40)$ & 49.8 & 34.3 & 1.4 (8)&  & 49.9 & 33.4 & 1.4 (8)\\ 
   & $(60,60,60,60,40)$ & 49.4 & 35.0 &  1.4 (8)&  & 49.6 & 34.1 & 1.4 (8)\\ 
	&  $(60,50,40,50,60)$ & 51.0 & 32.8 & 1.1 (6)&  & 50.9 & 32.5 & 1.1 (6)\\
		&   &  &  &&  & & &    \\

$(05,10, \boldsymbol{25},40,55)$ &  $(00,00,00,00,00)$&{\bf 57.7} & {\bf 21.5} &{\bf 0 (0)} &  &{\bf 57.7} &{\bf 21.5} &{\bf 0 (0)} \\  
   &  $(20,20,20,20,20)$ & 54.6 & 24.3 & 0.5 (3)&  & 54.5 & 24.1 &0.5 (3)\\ 
	 &  $(40,40,40,40,40)$ &53.1 & 25.7 &0.9 (5)&  & 53.1 & 25.5 &0.9 (5)\\ 
   & $(60,60,60,60,60)$ &51.2 & 27.8 &  1.4 (8)&  & 50.9 & 27.3 & 1.4 (8) \\ 
	 & $(80,80,80,80,80)$ & 49.4 & 29.6 & 2.0 (11)&  & 49.2 & 29.1 & 2.0 (11)\\ 
   & $(60,50,40,30,20)$ &54.3 & 24.4 &0.9 (5) &  & 54.4 & 23.9 &0.9 (5)\\ 
   &  $(60,40,40,40,40)$ &53.8 & 25.5 &0.9 (5)&  & 53.5 & 25.2 &0.9 (5)\\ 
   &  $(60,60,40,40,40)$ &52.4 & 25.6 &  1.1 (6)&  & 52.2 & 25.5 & 1.1 (6)\\ 
   &  $(60,60,60,40,40)$ & 53.5 & 25.5 & 1.3 (7)&  & 53.2 & 25.2 & 1.3 (7)\\ 
   & $(60,60,60,60,40)$ & 52.1 & 27.0 & 1.4 (8)&  & 51.7 & 26.6 & 1.4 (8)\\ 
	 &  $(60,50,40,50,60)$ &  52.6 & 26.6 & 1.1 (6)&  & 52.4 & 26.5 & 1.1 (6) \\
		&   &  &  &&  & & &    \\
	
$(01, 05,10, \boldsymbol{25},40)$ &   $(00,00,00,00,00)$ &{\bf 58.0} &{\bf 17.8} &{\bf 0 (0)} &  &{\bf 58.0} &{\bf 17.8 }&{\bf 0 (0)}\\ 
   &  $(20,20,20,20,20)$ & 56.1 & 19.5 &0.5 (3) &  & 55.8 & 19.4 &0.5 (3)\\ 
	   &  $(40,40,40,40,40)$ &54.3 & 20.8 & 0.9 (5)&  & 54.2 & 20.6 &0.9 (5)\\ 
   & $(60,60,60,60,60)$ & 51.9 & 22.8 & 1.4 (8)&  & 51.6 & 22.2 & 1.4 (8)\\ 
	   & $(80,80,80,80,80)$ &49.3 & 25.0 & 2.0 (11)&  & 48.8 & 24.4 & 2.0 (11)\\ 
   & $(60,50,40,30,20)$ &55.5 & 19.7 & 0.7 (4) &  & 55.5 & 19.4 &0.7 (4)\\ 
   &  $(60,40,40,40,40)$ &54.9 & 20.2 &0.9 (5) &  & 54.7 & 20.0 &0.9 (5)\\ 
   &  $(60,60,40,40,40)$ &54.2 & 20.8 &0.9 (5)&  & 54.1 & 20.6 &0.9 (5) \\ 
   &  $(60,60,60,40,40)$ &53.2 & 20.6 & 1.1 (6)&  & 52.7 & 20.2 & 1.1 (6)\\ 
   & $(60,60,60,60,40)$ &53.0 & 20.8 & 1.3 (7)&  & 53.1 & 20.3 & 1.3 (7)\\ 
	&  $(60,50,40,50,60)$ & 52.2 & 22.9 & 1.1 (6)&  & 52.5 & 22.3 & 1.1 (6)\\
		&   &  &  &&  & & &    \\

$(00, 01, 05,10, \boldsymbol{25})$ &   $(00,00,00,00,00)$ & {\bf 67.8} & n/a &{\bf 0 (0)} &  & {\bf 67.8} & n/a &{\bf 0 (0)}\\ 
   &  $(20,20,20,20,20)$ & 65.6 & n/a & 0.5 (3) &  & 65.5 & n/a &0.5 (3)\\ 
	   &  $(40,40,40,40,40)$ & 65.7 & n/a  &0.9 (5)&  & 65.3 & n/a &0.9 (5)\\ 
   & $(60,60,60,60,60)$ & 66.4 & n/a & 1.4 (8)&  & 65.7 &  n/a & 1.4 (8)\\ 
	   & $(80,80,80,80,80)$ & 65.9 & n/a & 2.0 (11)&  & 65.0 & n/a & 2.0 (11)\\ 
   & $(60,50,40,30,20)$ &65.5 & n/a & 0.7 (4)&  & 65.2 & n/a& 0.7 (4)\\ 
   &  $(60,40,40,40,40)$ & 65.6 & n/a &0.9 (5)&  & 65.2 & n/a &0.9 (5)\\ 
   &  $(60,60,40,40,40)$ &  65.6 & n/a &0.9 (5)&  & 65.2 & n/a &0.9 (5)\\ 
   &  $(60,60,60,40,40)$ & 66.0 &  n/a & 1.1 (6)&  & 65.7 &  n/a & 1.1 (6)\\ 
   & $(60,60,60,60,40)$ & 65.5 &  n/a & 1.3 (7)&  & 64.7 &  n/a & 1.3 (7)\\ 
	&  $(60,50,40,50,60)$ & 65.3 &  n/a & 1.3 (7)&  & 64.8 &  n/a & 1.3 (7)\\

\hline
\end{tabular}
}
\end{table}

\begin{table}[ht!]
\centering
{\small

\caption{Percent of correct selection (PCS), overdose selection (POS) and increase in sample size ($+N$ (\%)) for strategies A, B and C, according to dose limiting toxicity (DLT) and progression latent rates at $T$ (\%), in scenarios with 5 candidate dose levels, from 10000 simulations with $N=24$, $\phi=0.25$ and 2 patients-per-observation window accrual rate} \label{tab:res2425}
\begin{tabular}{llccccccc}
  \hline
    &&  \multicolumn{3}{c}{Strategy B}&&  \multicolumn{3}{c}{Strategy C}\\ \cline{3-5}\cline{7-9}
Pr(DLT)&Pr(Progression) &  PCS & POS & $+N$ (\%) &&  PCS &POS & $+N$ (\%) \\  \hline
	$(\boldsymbol{25},40,55,65,70)$ &  $(00,00,00,00,00)$ &{\bf 68.5}&{\bf 31.5} &{\bf 0 (0)}&  & {\bf 68.5 }&{\bf 31.5} &{\bf 0 (0)}\\ 
	 & $(20,20,20,20,20)$ & 65.2 & 34.8 & 0.7 (3) &  & 65.5 & 34.5 & 0.7 (3)\\ 
   & $(40,40,40,40,40)$ & 61.9 & 38.1 & 1.2 (5) &  & 62.4 & 37.6 & 1.2 (5)\\ 
  & $(60,60,60,60,60)$ & 57.8 & 42.2 &1.9 (8) &  & 59.2 & 40.8 &1.9 (8) \\ 
   & $(80,80,80,80,80)$  & 53.9 & 46.1 & 2.6 (11) &  & 55.8 & 44.2 & 2.6 (11)\\ 
   &  $(60,50,40,30,20)$ & 60.4 & 39.6 & 1.7 (7) &  & 61.0 & 39.0 &1.7 (7)\\ 
   & $(60,40,40,40,40)$ & 61.5 & 38.5 & 1.4 (6) &  & 62.0 & 38.0 &1.7 (7)\\ 
   & $(60,60,40,40,40)$  & 59.7 & 40.3 &1.7 (7)&  & 60.8 & 39.2 &1.7 (7)\\ 
   & $(60,60,60,40,40)$ & 58.2 & 41.8 &1.9 (8) &  & 59.5 & 40.5 &1.9 (8) \\ 
   & $(60,60,60,60,40)$ & 58.1 & 41.9 &1.9 (8) &  & 59.4 & 40.6 &1.9 (8) \\ 
   &  $(60,50,40,50,60)$ & 59.7 & 40.3 &1.7 (7)&  & 60.7 & 39.3 &1.7 (7)\\ 
		&   &  &  &&  & & &    \\

$(10, \boldsymbol{25},40,55,65)$  & $(00,00,00,00,00)$&{\bf 62.7 }&{\bf 25.0 }&{\bf 0 (0)}&  &{\bf 62.7} &{\bf 25.0} &{\bf 0 (0)}\\  
	 & $(20,20,20,20,20)$ & 61.0 & 27.4 &0.7 (3)&  & 61.1 & 27.3 &0.7 (3) \\ 
   & $(40,40,40,40,40)$  & 58.3 & 30.1 & 1.2 (5)&  & 58.2 & 29.7 & 1.2 (5)\\ 
   & $(60,60,60,60,60)$ & 55.2 & 33.3 &1.9 (8) &  & 55.3 & 32.5 &1.9 (8) \\ 
   &  $(80,80,80,80,80)$ & 51.7 & 36.7 & 2.6 (11) &  & 52.0 & 35.6 & 2.6 (11)\\ 
   &  $(60,50,40,30,20)$ & 56.9 & 31.2 & 1.4 (6)&  & 56.9 & 30.6 & 1.4 (6)\\ 
   &  $(60,40,40,40,40)$ & 57.4 & 30.6 & 1.4 (6)&  & 57.6 & 29.6 & 1.4 (6)\\ 
   & $(60,60,40,40,40)$  & 57.6 & 30.9 &1.7 (7)&  & 57.8 & 30.0 &1.7 (7)\\ 
   & $(60,60,60,40,40)$ & 55.8 & 32.9 &1.9 (8) &  & 56.0 & 32.0 &1.9 (8) \\ 
   & $(60,60,60,60,40)$ & 55.8 & 32.9 &1.9 (8) &  & 55.7 & 32.3 &1.9 (8) \\ 
   &  $(60,50,40,50,60)$  & 57.6 & 30.7 & 1.4 (6)&  & 57.5 & 30.1 & 1.4 (6)\\ 
		&   &  &  &&  & & &    \\

$(05,10, \boldsymbol{25},40,55)$ & $(00,00,00,00,00)$ &{\bf 64.6 }&{\bf 18.9 }&{\bf 0 (0)} &  &{\bf 64.6} &{\bf 18.9}&{\bf 0 (0)} \\ 
	 & $(20,20,20,20,20)$  & 62.7 & 20.6 &0.7 (3)&  & 62.2 & 20.7 &0.7 (3)\\ 
   & $(40,40,40,40,40)$ & 60.1 & 23.5 & 1.2 (5)&  & 60.2 & 23.0 & 1.2 (5)\\ 
   & $(60,60,60,60,60)$  & 57.8 & 25.7 &1.9 (8) &  & 57.8 & 24.7 &1.9 (8) \\ 
   & $(80,80,80,80,80)$ & 53.9 & 29.4 & 2.6 (11)&  & 54.5 & 27.8 & 2.6 (11)\\ 
   & $(60,50,40,30,20)$  & 61.7 & 21.4 & 1.2 (5)&  & 61.3 & 21.2 & 1.2 (5)\\ 
   &  $(60,40,40,40,40)$ & 60.1 & 23.7 & 1.2 (5)&  & 60.4 & 23.1 & 1.2 (5)\\ 
   &  $(60,60,40,40,40)$  & 59.7 & 23.2 & 1.4 (6)&  & 59.3 & 23.1 & 1.4 (6)\\ 
   & $(60,60,60,40,40)$  & 59.2 & 24.1 &1.7 (7)&  & 59.1 & 23.5 &1.7 (7)\\ 
   & $(60,60,60,60,40)$ & 57.6 & 25.8 & 1.9 (8) &  & 57.4 & 25.1 & 1.9 (8) \\ 
   & $(60,50,40,50,60)$ & 58.6 & 24.7 & 1.4 (6)&  & 58.4 & 24.3 &  1.4 (6)\\ 
	
		&   &  &  &&  & & &    \\
	
$(01, 05,10, \boldsymbol{25},40)$ & $(00,00,00,00,00)$&{\bf 65.5 }&{\bf 13.7}&{\bf 0 (0)} &&{\bf 65.5}&{\bf 13.7}&{\bf 0 (0)}\\  
 & $(20,20,20,20,20)$ & 62.1 & 16.2 &0.7 (3)&  & 62.0 & 16.0 & 0.7 (3)\\ 
   &  $(40,40,40,40,40)$ & 60.2 & 17.8 & 1.2 (5)&  & 60.0 & 17.5 & 1.2 (5)\\ 
   & $(60,60,60,60,60)$ & 58.8 & 19.3 &1.9 (8) &  & 58.5 & 18.6 &1.9 (8) \\ 
   & $(80,80,80,80,80)$ & 56.0 & 21.2 & 2.6 (11)&  & 56.0 & 20.1 & 2.6 (11)\\ 
   &  $(60,50,40,30,20)$  & 61.1 & 16.5 & 1.0 (4) &  & 60.9 & 16.2 & 1.0 (4)\\ 
   & $(60,40,40,40,40)$ & 60.3 & 17.9 & 1.2 (5)&  & 60.1 & 17.6 & 1.2 (5)\\ 
   & $(60,60,40,40,40)$ & 60.0 & 17.9 & 1.2 (5)&  & 59.8 & 17.5 & 1.2 (5)\\ 
   &  $(60,60,60,40,40)$  & 59.2 & 17.7 & 1.4 (6)&  & 58.9 & 17.5 & 1.4 (6)\\ 
  & $(60,60,60,60,40)$  & 60.0 & 18.2 &1.7 (7)&  & 59.3 & 17.9 &1.7 (7)\\ 
   & $(60,50,40,50,60)$  & 59.4 & 18.9 &  1.4 (6)&  & 59.0 & 18.5 &  1.4 (6)\\

		&   &  &  &&  & & &    \\

$(00, 01, 05,10, \boldsymbol{25})$ & $(00,00,00,00,00)$& {\bf 69.1} & n/a &{\bf 0 (0)}&&{\bf 69.1} & n/a &{\bf 0 (0)}\\  
	 &  $(20,20,20,20,20)$ & 69.8 & n/a  &0.7 (3) &  & 69.4 & n/a  &0.7 (3)\\ 
   &  $(40,40,40,40,40)$  & 68.7 & n/a  & 1.2 (5)&  & 68.4 & n/a  & 1.2 (5) \\ 
   & $(60,60,60,60,60)$ & 68.8 & n/a  &1.9 (8)  &  & 67.7 & n/a  & 1.9 (8) \\ 
   &  $(80,80,80,80,80)$ & 68.1 & n/a  & 2.6 (11)&  & 67.0 & n/a  & 2.6 (11)\\ 
   & $(60,50,40,30,20)$  & 69.2 & n/a  & 0.7 (3)&  & 68.9 & n/a  & 0.7 (3)\\ 
   & $(60,40,40,40,40)$ & 69.3 & n/a  & 1.2 (5)&  & 68.9 & n/a  & 1.2 (5)\\ 
   & $(60,60,40,40,40)$  & 69.3 & n/a  & 1.2 (5)&  & 69.1 &n/a  & 1.2 (5)\\ 
   & $(60,60,60,40,40)$ & 68.9 & n/a  & 1.4 (6)&  & 68.6 & n/a  &  1.4 (6)\\ 
   & $(60,60,60,60,40)$ & 68.5 & n/a  &1.7 (7)&  & 67.5 & n/a  &1.7 (7)\\ 
   &  $(60,50,40,50,60)$ & 69.3 & n/a  &1.7 (7) &  & 68.5 & n/a  & 1.7 (7)\\ 

\hline
\end{tabular}
}
\end{table}

\begin{table}[ht!]
\centering
{\small
\caption{Percent of correct selection (PCS), overdose selection (POS) and increase in sample size ($+N$ (\%)) for strategies A, B and C, according to dose limiting toxicity (DLT) and progression latent rates at $T$ (\%), in scenarios with 5 candidate dose levels, from 10000 simulations with $N=18$, $\phi=0.75$ and 2 patients-per-observation window accrual rate} \label{tab:res1875}
\begin{tabular}{llccccccc}
  \hline
    &&  \multicolumn{3}{c}{Strategy B}&&  \multicolumn{3}{c}{Strategy C}\\ \cline{3-5}\cline{7-9}
Pr(DLT)&Pr(Progression) &  PCS & POS & $+N$ (\%) &&  PCS &POS & $+N$ (\%) \\  \hline
$(\boldsymbol{25},40,55,65,70)$ &   $(00,00,00,00,00)$ & {\bf 66.0} & {\bf 34.0 }& {\bf 0 (0)} &  & {\bf 66.0 }& {\bf 34.0} & {\bf 0 (0)} \\ 
   &  $(20,20,20,20,20)$ &65.3 & 34.7 & 2.3 (13) &  & 67.9 & 32.2 &2.3 (13)\\ 
	   &  $(40,40,40,40,40)$ &  64.0 & 36.0 & 5.0 (28) &  & 69.9 & 30.1 & 5.2 (29) \\ 
   & $(60,60,60,60,60)$ &62.5 & 37.5 & 8.8 (49) &  & 71.9 & 28.1 & 9.2 (51)\\ 
	   & $(80,80,80,80,80)$ & 59.9 & 40.1 & 14.2 (79) &  & 75.7 & 24.3 & 14.8 (82)  \\ 
   & $(60,50,40,30,20)$ &63.3 & 36.7 & 7.6 (42) &  & 71.3 & 28.7 & 7.9 (44) \\ 
   &  $(60,40,40,40,40)$ &63.5 & 36.5 & 7.0 (39) &  & 70.9 & 29.1 & 7.4 (41)\\ 
   &  $(60,60,40,40,40)$ & 63.3 & 36.7 & 8.1 (45) &  & 72.2 & 27.8 & 8.5 (47) \\ 
   &  $(60,60,60,40,40)$ & 62.5 & 37.5 &8.8 (49) &  & 72.0 & 28.0 & 9.0 (50) \\ 
   & $(60,60,60,60,40)$ &62.4 & 37.7 &8.8 (49) &  & 72.2& 27.8 &9.2 (51)\\ 
	&  $(60,50,40,50,60)$ &63.1 & 36.9 & 7.7 (43) &  & 71.7 & 28.3 & 7.9 (44)\\
		&   &  &  &&  & & &    \\

$(10, \boldsymbol{25},40,55,65)$  &  $(00,00,00,00,00)$&{\bf 57.6 }&{\bf 27.0 }&{\bf 0 (0)}&  &{\bf 57.6 }& {\bf27.0} &{\bf 0 (0)}\\  
   &  $(20,20,20,20,20)$ &56.3 & 28.3 &2.3 (13)&  & 57.2 & 26.2 &2.3 (13) \\ 
	   &  $(40,40,40,40,40)$ & 56.0 & 29.7 &5.2 (29)&  & 57.1 & 24.7 & 5.4 (30) \\ 
   & $(60,60,60,60,60)$ &55.5 & 31.3 & 9.4 (52) &  & 57.2 & 22.2 & 9.5 (53)\\ 
	   & $(80,80,80,80,80)$ &54.5 & 33.7 & 14.8 (82) &  & 56.3 & 19.6 & 15.5 (86)\\ 
   & $(60,50,40,30,20)$ & 55.6 & 30.9 & 6.7 (37) &  & 56.1 & 24.6 &7.0 (39)\\ 
   &  $(60,40,40,40,40)$ & 56.4 & 29.9 & 6.1 (34) &  & 55.8 & 25.2 & 6.3 (35)\\ 
   &  $(60,60,40,40,40)$ & 56.1 & 31.2 & 7.7 (43) &  & 55.7 & 24.4 &8.1 (45)\\ 
   &  $(60,60,60,40,40)$ & 55.7 & 31.1 &8.8 (49)&& 56.5 & 22.9 &9.2 (51)\\ 
   & $(60,60,60,60,40)$ & 55.2 & 31.6 &9.2 (51)&& 56.9 & 22.4 & 9.5 (53)\\ 
	&  $(60,50,40,50,60)$ & 55.9 & 30.6 & 7.0 (39) && 56.5 & 24.2 &7.4 (41)\\
		&   &  &  &&  & & &    \\

$(05,10, \boldsymbol{25},40,55)$ &  $(00,00,00,00,00)$&{\bf 57.7} & {\bf21.5 }&{\bf 0 (0)}&  &{\bf 57.7} & {\bf21.5} & {\bf 0 (0)}\\  
     &  $(20,20,20,20,20)$ &57.0 & 23.4 &2.3 (13)&  & 56.9 & 21.3 &2.3 (13)\\ 
	   &  $(40,40,40,40,40)$ & 57.8 & 24.1 &5.4 (30)&  & 57.1 & 19.9 &5.4 (30)\\ 
   & $(60,60,60,60,60)$ &57.5 & 25.6 & 9.5 (53)&  & 56.1 & 18.4 & 9.7 (54)\\ 
	   & $(80,80,80,80,80)$ &57.0 & 27.9 & 15.1 (84) &  & 56.0 & 15.5 & 15.8 (88)\\ 
   & $(60,50,40,30,20)$ &58.9 & 23.6 & 5.6 (31) &  & 56.69 & 20.0 & 5.8 (32) \\ 
   &  $(60,40,40,40,40)$ &57.7 & 24.3 &5.6 (31)&  & 56.8 & 19.9 &5.8 (32)\\ 
   &  $(60,60,40,40,40)$ & 57.5 & 24.8 & 6.5 (36) &  & 56.3 & 19.7 & 6.8 (38)\\ 
   &  $(60,60,60,40,40)$ &58.3 & 24.3 & 8.3 (46) &  & 55.8 & 18.1 &8.8 (49)\\ 
   & $(60,60,60,60,40)$ &58.0 & 25.4&9.2 (51)&  & 55.7 & 18.5 & 6.5 (53)\\ 
	&  $(60,50,40,50,60)$ & 58.0 & 24.7 & 6.7 (37) &  & 58.1 & 19.0 & 6.8 (38)  \\
		&   &  &  &&  & & &    \\

$(01, 05,10, \boldsymbol{25},40)$ &   $(00,00,00,00,00)$ &{\bf 58.0} & {\bf17.8} & {\bf 0 (0)}&  & {\bf58.0}  &{\bf 17.8}& {\bf 0 (0)}\\ 
   &  $(20,20,20,20,20)$ & 57.7 & 19.1 &2.3 (13)&  & 57.4 & 17.1 & 2.3 (13)\\ 
	   &  $(40,40,40,40,40)$ &58.2 & 19.8 &5.4 (30)&  & 57 & 16.1 &  5.6 (31)\\ 
   & $(60,60,60,60,60)$ &  58.6 & 21.2 & 9.7 (54)&  & 55.2 & 14.8 & 9.9 (55) \\ 
	   & $(80,80,80,80,80)$ &58.9 & 22.1 & 15.7 (87) &  & 54.0 & 11.9 & 16.2 (90) \\ 
   & $(60,50,40,30,20)$ &59.1 & 19.4 & 4.3 (24) &  & 57.0 & 16.8 & 4.5 (25)\\ 
   &  $(60,40,40,40,40)$ &58.8 & 19.4 & 5.4 (30) &  & 57.0 & 15.9 &5.6 (31) \\ 
   &  $(60,60,40,40,40)$ & 58.6 & 19.4 & 5.8 (32)&  & 56.6 & 16.2 & 5.9 (33) \\ 
   &  $(60,60,60,40,40)$ & 59.8 & 19.3 &7.0 (39)&  & 56.5 & 15.9 &7.4 (41)\\ 
   & $(60,60,60,60,40)$ & 58.9 & 20.0 &8.8 (49)&  & 55.5 & 14.8 &9.2 (51)\\ 
	&  $(60,50,40,50,60)$ & 57.5 & 20.9 & 7.2 (40) &  & 57.5 & 14.8 & 7.2 (40)\\
		&   &  &  &&  & & &    \\

$(00, 01, 05,10, \boldsymbol{25})$ &   $(00,00,00,00,00)$ & {\bf 67.8} & n/a &{\bf 0 (0)}&  &{\bf 67.8} & n/a &{\bf 0 (0)}\\ 
   &  $(20,20,20,20,20)$ & 68.8 & n/a&2.3 (13)&  & 66.1 & n/a & 2.5 (14)\\ 
	   &  $(40,40,40,40,40)$ & 70.8 &  n/a & 5.6 (31)&  & 65.1 &  n/a & 5.8 (32)\\ 
   & $(60,60,60,60,60)$ &72.9 & n/a & 9.9 (55)&  & 63.4 &n/a & 10.1 (56)\\ 
	   & $(80,80,80,80,80)$ & 75.0 & n/a & 16.0 (89) &  & 58.7 &n/a& 16.7 (93) \\ 
   & $(60,50,40,30,20)$ & 69.9 & n/a & 3.6 (20) &  & 66.5 & n/a & 3.6 (20)\\ 
   &  $(60,40,40,40,40)$ & 70.9 & n/a &5.6 (31)&  & 64.7 & n/a &5.8 (32)\\ 
   &  $(60,60,40,40,40)$ & 70.7 &  n/a &5.6 (31)&  & 65.0 &  n/a &5.8 (32)\\ 
   &  $(60,60,60,40,40)$ & 71.6 & n/a & 6.3 (35) &  & 65.2 &  n/a & 6.5 (36)\\ 
   & $(60,60,60,60,40)$ & 72.5 & n/a &7.7 (43)&  & 64.2 & n/a &8.1 (45)\\ 
	&  $(60,50,40,50,60)$ &71.4 & n/a &8.5 (47)&  & 63.5 & n/a & 8.5 (47)\\

\hline
\end{tabular}
}
\end{table}

\begin{table}[ht!]
\centering
{\small
\caption{Percent of correct selection (PCS), overdose selection (POS) and increase in sample size ($+N$ (\%)) for strategies A, B and C, according to dose limiting toxicity (DLT) and progression latent rates at $T$ (\%), in scenarios with 5 candidate dose levels, from 10000 simulations with $N=24$, $\phi=0.75$ and 2 patients-per-observation window accrual rate} \label{tab:res2475}
\begin{tabular}{llccccccc}
  \hline
    &&  \multicolumn{3}{c}{Strategy B}&&  \multicolumn{3}{c}{Strategy C}\\ \cline{3-5}\cline{7-9}
	 Pr(DLT)&Pr(Progression) &  PCS & POS & $+N$ (\%) &&  PCS &POS & $+N$ (\%) \\  \hline
	$(\boldsymbol{25},40,55,65,70)$ &  $(00,00,00,00,00)$ &{\bf 68.5}&{\bf 31.5} &{\bf 0 (0)}&  & {\bf 68.5 }&{\bf 31.5} &{\bf 0 (0)}\\ 
	 & $(20,20,20,20,20)$  & 67.2 & 32.8 & 2.9 (12) &  & 69.9 & 30.1 & 3.1 (13)\\ 
   &  $(40,40,40,40,40)$ & 65.0& 35.0 & 6.7 (28) &  & 71.9 & 28.1 & 7.0 (29) \\ 
   & $(60,60,60,60,60)$  & 62.6 & 37.4 & 12.0 (50) &  & 74.4 & 25.6 & 12.2 (51) \\ 
   & $(80,80,80,80,80)$  & 59.8 & 40.2 & 19.2 (80) &  & 78.2 & 21.8 & 19.7 (82) \\ 
   &  $(60,50,40,30,20)$ & 64.1 & 35.9 & 10.1 (42) &  & 73.0 & 27.0 & 10.6 (44) \\ 
   & $(60,40,40,40,40)$  & 64.7 & 35.3 & 9.4 (39) &  & 72.7 & 27.3 &10.1 (42) \\ 
   & $(60,60,40,40,40)$ & 62.9 & 37.1 & 11.3 (47) &  & 73.3 & 26.7 & 11.5 (48) \\ 
   &  $(60,60,60,40,40)$ & 62.6 & 37.4 & 11.8 (49) &  & 74.1 & 25.9 & 12.2 (51) \\ 
   & $(60,60,60,60,40)$ & 62.5 & 37.5 &12.0 (50)&  & 74.2 & 25.9 & 12.2 (51)\\ 
   &  $(60,50,40,50,60)$  & 64.2 & 35.8 & 10.3 (43) &  & 73.3 & 26.7 & 10.8 (45) \\ 
		&   &  &  &&  & & &    \\

$(10, \boldsymbol{25},40,55,65)$  & $(00,00,00,00,00)$&{\bf 62.7 }&{\bf 25.0 }&{\bf 0 (0)}&  &{\bf 62.7} &{\bf 25.0} &{\bf 0 (0)}\\  
	 &  $(20,20,20,20,20)$ & 62.1 & 26.6 &3.1 (13)&  & 63.0 & 24.0 &3.1 (13)\\ 
   &  $(40,40,40,40,40)$  & 62.2 & 27.7 &7.0 (29)&  & 64.0 & 21.6 & 7.2 (30) \\ 
   & $(60,60,60,60,60)$  & 61.0 & 29.3 &12.2 (51)&  & 64.0 & 19.4 & 12.7 (53) \\ 
   & $(80,80,80,80,80)$ & 60.2 & 31.6 &19.7 (82)&  & 64.1 & 16.0 & 20.6 (86)\\ 
   & $(60,50,40,30,20)$ & 62.2 & 28.3 & 8.9 (37) &  & 63.3 & 21.7 & 9.4 (39)\\ 
   & $(60,40,40,40,40)$ & 62.7 & 27.6 & 7.7 (32) &  & 63.9 & 21.7 & 8.2 (34) \\ 
   & $(60,60,40,40,40)$ & 62.1 & 28.0 &10.3 (43)&  & 62.6 & 20.8 & 10.8 (45)\\ 
   & $(60,60,60,40,40)$ & 61.0 & 29.1 & 12.0 (50) &  & 63.6 & 20.0 &12.2 (51)\\ 
   & $(60,60,60,60,40)$ & 61.1 & 29.0 & 12.2 (51)&  & 64.1 & 19.2 & 12.5 (52) \\ 
   & $(60,50,40,50,60)$ & 62.2 & 28.1 & 9.1 (38) &  & 63.2 & 21.4 & 9.6 (40) \\ 
		&   &  &  &&  & & &    \\

$(05,10, \boldsymbol{25},40,55)$ & $(00,00,00,00,00)$ &{\bf 64.6 }&{\bf 18.9 }&{\bf 0 (0)} &  &{\bf 64.6} &{\bf 18.9}&{\bf 0 (0)} \\ 
	 &  $(20,20,20,20,20)$ & 64.3 & 20.0 &3.1 (13)&  & 63.8 & 18.1 &3.1 (13)\\ 
   &  $(40,40,40,40,40)$ & 64.6 & 21.1 & 7.2 (30)&  & 63.8 & 16.6 &7.2 (30)\\ 
   & $(60,60,60,60,60)$ & 64.7 & 22.7 & 12.5 (52)&  & 63.9 & 14.7 & 13.0 (54) \\ 
   & $(80,80,80,80,80)$ & 63.0 & 25.3 & 20.4 (85) &  & 62.4 & 11.9 & 21.1 (88) \\ 
   & $(60,50,40,30,20)$ & 65.0 & 20.9 & 7.2 (30)&  & 62.9 & 17.4 & 7.4 (31) \\ 
   & $(60,40,40,40,40)$ & 64.6 & 21.0 & 7.2 (30) &  & 63.6 & 16.8 & 7.4 (31)\\ 
   & $(60,60,40,40,40)$ & 64.9 & 21.8 & 8.4 (35) &  & 63.2 & 16.9 &8.9 (37)\\ 
   & $(60,60,60,40,40)$ & 65.2 & 22.1 &11.3 (47)&  & 63.1 & 15.7 &11.8 (49)\\ 
   &  $(60,60,60,60,40)$  & 64.9 & 22.2 & 12.5 (52) &  & 63.9 & 14.8 &12.7 (53)\\ 
   &  $(60,50,40,50,60)$  & 63.8 & 22 & 8.6 (36) &  & 64.5 & 16.0 & 8.6 (36)\\

		&   &  &  &&  & & &    \\
	
$(01, 05,10, \boldsymbol{25},40)$ & $(00,00,00,00,00)$&{\bf 65.5 }&{\bf 13.7}&{\bf 0 (0)} &&{\bf 65.5}&{\bf 13.7}&{\bf 0 (0)}\\  
 & $(20,20,20,20,20)$ &  64.9 & 15.3 &3.1 (13)&  & 64.7 & 12.9 &3.1 (13)\\ 
   &  $(40,40,40,40,40)$ & 65.5 & 16.1 &7.2 (30)&  & 64.5 & 11.7 & 7.4 (31)\\ 
   & $(60,60,60,60,60)$ & 65.7 & 17.3 & 13.0 (54) &  & 62.3 & 11.1 & 13.2 (55)\\ 
   & $(80,80,80,80,80)$ & 65.7 & 19.0 &20.6 (86)&  & 60.0 & 8.8 & 21.4 (89) \\ 
   &  $(60,50,40,30,20)$  &  66.5 & 15.1 & 5.5 (23) &  & 63.6 & 12.8 & 5.8 (24)\\ 
   & $(60,40,40,40,40)$ & 65.3 & 16.2 &7.2 (30) &  & 64.0 & 11.9 & 7.4 (31)\\ 
   & $(60,60,40,40,40)$ &  65.1 & 15.9 & 7.4 (31)&  & 63.5 & 12.2 &7.7 (32)\\ 
   &  $(60,60,60,40,40)$  &  66.0 & 16.0& 9.1 (38) &  & 62.9 & 11.9 &9.6 (40)\\ 
  & $(60,60,60,60,40)$  & 66.2 & 17.0 & 11.8 (49) &  & 62.1 & 11.2 &12.2 (51)\\ 
   & $(60,50,40,50,60)$  &  64.4 & 17.1 &9.6 (40)&  & 64.0 & 11.1 &9.6 (40)\\ 
	
		&   &  &  &&  & & &    \\

$(00, 01, 05,10, \boldsymbol{25})$ & $(00,00,00,00,00)$& {\bf 69.1} & n/a &{\bf 0 (0)}&&{\bf 69.1} & n/a &{\bf 0 (0)}\\  
& $(20,20,20,20,20)$ & 71.3 & n/a &3.1 (13)&  & 68.6 & n/a & 3.4 (14) \\ 
   & $(40,40,40,40,40)$ & 73.5 & n/a & 7.4 (31)&  & 67.3 & n/a& 7.7 (32) \\ 
   & $(60,60,60,60,60)$ & 75.6 & n/a & 13.2 (55) &  & 64.5 & n/a &13.2 (55) \\ 
   & $(80,80,80,80,80)$  & 78.3 & n/a & 21.4 (89) &  & 60.8 & n/a & 22.1 (92) \\ 
   & $(60,50,40,30,20)$ & 72.6 & n/a & 4.6 (19) &  & 68.3 & n/a & 4.6 (19) \\ 
   & $(60,40,40,40,40)$ & 73.4 & n/a & 7.4 (31)&  & 66.9 & n/a & 7.4 (31)\\ 
   &  $(60,60,40,40,40)$ & 73.5 & n/a & 7.4 (31)&  & 67.2 & n/a &7.7 (32)\\ 
   & $(60,60,60,40,40)$ & 74.0 & n/a & 8.2 (34) &  & 67.6 & n/a & 8.4 (35) \\ 
   & $(60,60,60,60,40)$  & 74.6 & n/a & 10.1 (42) &  & 65.9 & n/a & 10.6 (44) \\ 
   & $(60,50,40,50,60)$  & 74.1 & n/a & 11.3 (47)&  & 65.0 & n/a &11.3 (47)\\ 
\hline
\end{tabular}
}
\end{table}

\end{document}